# Variable Forgetting in Reasoning about Knowledge

**Kaile Su**                                                                    SUKL@PKU.EDU.CN
*School of Electronics Engineering and Computer Science*
*Peking University*
*Beijing, P.R. China*

**Abdul Sattar**                                                          A.SATTAR@GRIFFITH.EDU.AU
*Institute for IIS*
*Griffith University*
*Brisbane, Qld 4111, Australia*

**Guanfeng Lv**                                                                 LVGF@YAHOO.COM
*School of Computer Science*
*Beijing University of Technology*
*Beijing, P.R. China*

**Yan Zhang**                                                             YAN@CIT.UWS.EDU.AU
*School of Computing and Information Technology*
*University of Western Sydney*
*Penrith South DC NSW 1797, Australia*

## Abstract

In this paper, we investigate knowledge reasoning within a simple framework called *knowledge structure*. We use *variable forgetting* as a basic operation for one agent to reason about its own or other agents' knowledge. In our framework, two notions namely agents' *observable variables* and the *weakest sufficient condition* play important roles in knowledge reasoning. Given a background knowledge base $\Gamma$ and a set of observable variables $O_i$ for each agent $i$, we show that the notion of agent $i$ knowing a formula $\varphi$ can be defined as a weakest sufficient condition of $\varphi$ over $O_i$ under $\Gamma$. Moreover, we show how to capture the notion of common knowledge by using a generalized notion of weakest sufficient condition. Also, we show that public announcement operator can be conveniently dealt with via our notion of knowledge structure. Further, we explore the computational complexity of the problem whether an epistemic formula is *realized* in a knowledge structure. In the general case, this problem is PSPACE-hard; however, for some interesting subcases, it can be reduced to co-NP. Finally, we discuss possible applications of our framework in some interesting domains such as the automated analysis of the well-known muddy children puzzle and the verification of the revised Needham-Schroeder protocol. We believe that there are many scenarios where the natural presentation of the available information about knowledge is under the form of a knowledge structure. What makes it valuable compared with the corresponding multi-agent S5 Kripke structure is that it can be much more succinct.

## 1. Introduction

Epistemic logics, or logics of knowledge are usually recognized as having originated in the work of Jaakko Hintikka - a philosopher who showed how certain modal logics could be used to capture intuitions about the nature of knowledge in the early 1960s (Hintikka, 1962). In the mid of 1980s, Halpern and his colleagues discovered that S5 epistemic logics could be given a natural interpretation in terms of the states of processes (commonly called agents) in a distributed system. This model now is known as the *interpreted system* model (Fagin, Halpern, Moses, & Vardi, 1995). It was found that this model plays an important role in the theory of distributed systems and has been applied successfully in reasoning about communication protocols (Halpern & Zuck, 1992). However, the work on epistemic logic has mainly focused on theoretical issues such as variants of modal logic, completeness, computational complexity, and derived notions like distributed knowledge and common knowledge.

In this paper, we explore knowledge reasoning within a more concrete model of knowledge. Our framework of reasoning about knowledge is simple and powerful enough to analyze realistic protocols such as some widely used security protocols.

To illustrate the problem investigated in this paper, let us consider the communication scenario that Alice sends Bob a message and Bob sends Alice an acknowledgement when receiving the message. We assume Alice and Bob commonly have the following background knowledge base $\Gamma_{CS}$:

$$Bob\_recv\_msg \Rightarrow Alice\_send\_msg$$
$$Bob\_send\_ack \Rightarrow Bob\_recv\_msg$$
$$Alice\_recv\_ack \Rightarrow Bob\_send\_ack$$

where $Bob\_recv\_msg$ and $Bob\_send\_ack$ are *observable* variables to Bob, while $Alice\_send\_msg$ and $Alice\_recv\_ack$ are *observable* to Alice.

The problem we are concerned with is how to verify that Alice or Bob knows a statement $\varphi$. Intuitively, we should be able to prove that for a statement observable to Alice (Bob), Alice (Bob) knows the statement if and only if the statement itself holds. As for the knowledge of non-observable statements, the following should hold:

1. Alice knows $Bob\_recv\_msg$ if $Alice\_recv\_ack$ holds; on the other hand, if Alice knows $Bob\_recv\_msg$, then $Alice\_recv\_ack$ holds, which means that, in the context of this example, the only way that Alice gets to know $Bob\_recv\_msg$ is that Alice receives the acknowledgement from Bob.

2. Bob knows $Alice\_send\_msg$ if $Bob\_recv\_msg$ holds; moreover, if Bob knows $Alice\_send\_msg$, then $Bob\_recv\_msg$ holds. The latter indicates that the only way that Bob gets to know $Alice\_send\_msg$ is that Bob receives the message from Alice.

3. Finally, Bob does not know $Alice\_recv\_ack$.

The idea behind the presented knowledge model for those scenarios demonstrated above is that an agent's knowledge is just the agent's observations or logical consequences of the agent's observations under the background knowledge base.

One of the key notions introduced in this paper is agents' *observable variables*. This notion shares a similar spirit of those of *local variables* in the work of van der Hoek and Wooldridge (2002) and *local propositions* in the work of Engelhardt, van der Meyden and Moses (1998) and in the work of Engelhardt, van der Meyden and Su (2003). Informally speaking, local propositions are those depending only upon an agent's local information; and an agent can always determine whether a given local proposition is true. Local variables are those primitive propositions that are local. Nevertheless, the notion of local propositions (Engelhardt et al., 1998, 2003) is a semantics property of the truth assignment function in a Kripke structure, while the notion of local variables (van der Hoek & Wooldridge, 2002) is a property of syntactical variables. In this paper, we prefer to use the term "observable variable" in order to avoid any confusion with the term "local variable" used in programming, where "non-local variables" such as "global variables" may often be observable.

Our knowledge model is also closely related to the notion of *weakest sufficient condition*, which was first formalized by Lin (2001). Given a background knowledge base $\Gamma$ and a set of observable variables $O_i$ for each agent $i$, we show that the notion of agent $i$ knowing a formula $\varphi$ can be defined as the weakest sufficient condition of $\varphi$ over $O_i$ under $\Gamma$, which can be computed via the operation of *variable forgetting* (Lin & Reiter, 1994) or *eliminations of middle terms* (Boole, 1854). Moreover, we generalize the notion of weakest sufficient condition and capture the notion of common knowledge.

Now we briefly discuss the role of variable forgetting in our knowledge model. Let us examine the scenario described above again. Consider the question: how can Alice figure out Bob's knowledge when she receives the acknowledgement from Bob? Note that Alice's knowledge is the conjunction of the background knowledge base $\Gamma_{CS}$ and her observations $Alice\_recv\_ack$ etc. Moreover, all Alice knows about Bob's knowledge is the conjunction of the background knowledge base $\Gamma_{CS}$ and all she knows about Bob's observations. Thus, Alice gets Bob's knowledge by computing all she knows about Bob's observations. In our setting, Alice gets her knowledge on Bob's observations simply by forgetting Bob's non-observable variables in her own knowledge.

There is a recent trend of extending epistemic logics with dynamic operators so that the evolution of knowledge can be expressed (van Benthem, 2001; van Ditmarsch, van der Hoek, & Kooi, 2005a). The most basic extension is public announcement logic (PAL), which is obtained by adding an operator for truthful public announcements (Plaza, 1989; Baltag, Moss, & Solecki, 1998; van Ditmarsch, van der Hoek, & Kooi, 2005b). We show that public announcement operator can be conveniently dealt with via our notion of knowledge structure. This makes the notion of knowledge structure genuinely useful for those applications like the automated analysis of the well-known muddy children puzzle.

From the discussion above, we can see that our framework of reasoning about knowledge is appropriate in those situations where every agent has a specified set of observable variables. To further show the significance of our framework, we investigate some of its interesting applications to the automated analysis of the well-known muddy children puzzle and the verification of the revised Needham-Schroeder protocol (Lowe, 1996).

We believe that there are many scenarios where the natural presentation of the available information about knowledge is under the form of a knowledge structure. What makes it valuable compared with the corresponding multi-agent S5 Kripke structure is that it can be much more succinct. Of course, the price to pay is that determining whether a formula holds in a knowledge structure is PSPACE-hard in the general case, while it is in PTIME when the corresponding S5 Kripke structure is taken as input. However, the achieved trade-off between time and space can prove computationally valuable. In particular, the validity problem from a knowledge structure can be addressed for some instances for which generating the corresponding Kripke structure would be unfeasible. The muddy children puzzle shows this point clearly: generating the corresponding Kripke structure is impossible from a practical point of view, even for the least number of children considered in the experiments.

The organization of this paper is as follows. In the next section, we briefly introduce the concept of forgetting and the notion of weakest sufficient and strongest necessary conditions. In Section 3, we define our framework of reasoning about knowledge via variable forgetting. In Section 4, we generalize the notion of weakest sufficient condition and strongest necessary condition to capture common knowledge within our framework. In Section 5, we show that public announcement operator can also be conveniently dealt with via our notion of knowledge structure. Section 6 discusses the computational complexity issue about the problem of whether an epistemic formula is realized in a knowledge structure. In the general case, this problem is PSPACE-hard; however, for some interesting subcases, it can be reduced to co-NP. In Section 7, we consider a case study by applying our framework to model the well known muddy children puzzle; and further more to security protocol verification in Section 8. Finally, we discuss some related work and conclude the paper with some remarks.

## 2. Preliminaries

In this section, we provide some preliminaries about the notions of variable forgetting and weakest sufficient condition, and epistemic logic.

### 2.1 Forgetting

Given a set of propositional variables $P$, we identify a *truth assignment over $P$* with a subset of $P$. We say a formula $\varphi$ is a formula *over $P$* if each propositional variable occurring in $\varphi$ is in $P$. For convenience, we define **true** as an abbreviation for a fixed valid propositional formula, say $p \vee \neg p$, where $p$ is primitive proposition in $P$. We abbreviate $\neg$**true** by **false**.

We also use $\models$ to denote the usual satisfaction relation between a truth assignment and a formula. Moreover, for a set of formulas $\Gamma$ and a formula $\varphi$, we use $\Gamma \models \varphi$ to denote that for every assignment $\sigma$, if $\sigma \models \alpha$ for all $\alpha \in \Gamma$, then $\sigma \models \varphi$.

Given a propositional formula $\varphi$, and a propositional variable $p$, we denote by $\varphi(\frac{p}{\textbf{true}})$ the result of replacing every $p$ in $\varphi$ by **true**. We define $\varphi(\frac{p}{\textbf{false}})$ similarly.

The notion of *variable forgetting* (Lin & Reiter, 1994), or eliminations of middle terms (Boole, 1854), can be defined as follows:

**Definition 1** Let $\varphi$ be a formula over $P$, and $V \subseteq P$. The *forgetting of $V$ in $\varphi$*, denoted as $\exists V \varphi$, is a quantified formula over $P$, defined inductively as follows:

1. $\exists \emptyset \varphi = \varphi$;

2. $\exists \{p\} \varphi = \varphi \left( \frac{p}{\textbf{true}} \right) \vee \varphi \left( \frac{p}{\textbf{false}} \right)$;

3. $\exists (V \cup \{p\}) \varphi = \exists V (\exists \{p\} \varphi)$.

For convenience, we use $\forall V \varphi$ to denote $\neg \exists V (\neg \varphi)$.

**Example 2:** Let $\varphi = (p \vee q) \wedge (\neg p \vee r)$. We have $\exists \{p\} \varphi \equiv (q \vee r)$ and $\exists \{q\} \varphi \equiv (\neg p \vee r)$. □

Many characterizations of variable forgetting, together with complexity results, are reported in the work of Lang and Marquis (1998). In particular, the notion of variable forgetting is closely related to that of *formula-variable independence* (Lang, Liberatore, & Marquis, 2003).

**Definition 3** Let $\varphi$ be a propositional formula, and $V$ a set of propositional variables. We say $\varphi$ is independent from $V$ if and only if $\varphi$ is logically equivalent to a formula in which none of the variables in $V$ appears.

The following proposition was given in the work of Lang, Liberatore and Marquis (2003).

**Proposition 4** *Let $\varphi$ be a propositional formula, and $V$ a set of propositional variables. Then $\exists V \varphi$ is the logically strongest consequence of $\varphi$ that is independent from $V$ (up to logical equivalence).*

## 2.2 Weakest Sufficient Conditions

The formal definitions of *weakest sufficient conditions* and *strongest necessary conditions* were first formalized via the notion of variable forgetting by Lin (2001), which in turn play an essential role in our approach.

**Definition 5** Let $V$ be a set of propositional variables and $V' \subseteq V$. Given a set of formulas $\Gamma$ over $V$ as a background knowledge base and a formula $\alpha$ over $V$.

- A formula $\varphi$ over $V'$ is called a *sufficient condition of $\alpha$ over $V'$* under $\Gamma$ if $\Gamma \models \varphi \Rightarrow \alpha$. It is called a *weakest sufficient condition of $\alpha$ over $V'$* under $\Gamma$ if it is a sufficient condition of $\alpha$ over $V'$ under $\Gamma$, and for any sufficient condition $\varphi'$ of $\alpha$ on $V'$ under $\Gamma$, we have $\Gamma \models \varphi' \Rightarrow \varphi$.

- A formula $\varphi$ over $V'$ is called a necessary condition of $\alpha$ over $V'$ under $\Gamma$ if $\Gamma \models \alpha \Rightarrow \varphi$. It is called a *strongest necessary condition of $\alpha$ over $V'$* under $\Gamma$ if it is a necessary condition of $\alpha$ over $V'$ under $\Gamma$, and for any necessary condition $\varphi'$ of $\alpha$ over $V'$ under $\Gamma$, we have $\Gamma \models \varphi \Rightarrow \varphi'$.

The notions given above are closely related to theory of abduction. Given an observation, there may be more than one abduction conclusion that we can draw. It should be useful to find the weakest one of such conclusions, i.e., the weakest sufficient condition of the observation (Lin, 2001). The notions of strongest necessary and weakest sufficient conditions of a proposition also have many potential applications in other areas such as reasoning about actions. The following proposition, which is due to Lin (2001), shows how to compute the two conditions.

**Proposition 6** *Given a background knowledge base $\{\theta\}$ over $V$, a formula $\alpha$ over $V$, and a subset $V'$ of $V$. Let $SNC^\alpha$ and $WSC^\alpha$ be a strongest necessary condition and a weakest sufficient condition of $\alpha$ over $V'$ under $\{\theta\}$ respectively. Then*

- $WSC^\alpha$ *is equivalent to* $\forall (V - V')(\theta \Rightarrow \alpha)$; *and*

- $SNC^\alpha$ *is equivalent to* $\exists (V - V')(\theta \wedge \alpha)$.

### 2.3 Epistemic Logic and Kripke Structure

We now recall some standard concepts and notations related to the modal logics for multi-agents' knowledge.

Given a set $V$ of propositional variables. Let $\mathcal{L}(V)$ be the set of all propositional formulas on $V$. The language of epistemic logic, denoted by $\mathcal{L}_n(V)$, is $\mathcal{L}(V)$ augmented with modal operator $K_i$ for each agent $i$. $K_i\phi$ can be read *"agent $i$ knows $\phi$"*. Let $\mathcal{L}_n^C(V)$ be the language of $\mathcal{L}_n(V)$ augmented with modal operator $C_\Delta$ for each set of agents $\Delta$. A formula $C_\Delta\alpha$ indicates that it is common knowledge among agents in $\Delta$ that $\alpha$ holds. We omit the argument $V$ and write $\mathcal{L}_n$ and $\mathcal{L}_n^C$, if it is clear from context.

According to the paper by Halpern and Moses (1992), semantics of these formulas can be given by means of *Kripke structure* (Kripke, 1963), which formalizes the intuition behind possible worlds. A Kripke structure is a tuple $(W, \pi, \mathcal{K}_1, \cdots, \mathcal{K}_n)$, where $W$ is a set of *worlds*, $\pi$ associates with each world a truth assignment to the propositional variables, so that $\pi(w)(p) \in \{\mathbf{true}, \mathbf{false}\}$ for each world $w$ and propositional variable $p$, and $\mathcal{K}_1, \cdots, \mathcal{K}_n$ are binary accessibility relations. By convention, $W^M$, $\mathcal{K}_i^M$ and $\pi^M$ are used to refer to the set $W$ of possible worlds, the $\mathcal{K}_i$ relation and the $\pi$ function in the Kripke structure $M$, respectively. We omit the superscript $M$ if it is clear from context. Finally, let $\mathcal{C}_\Delta$ be the transitive closure of $\bigcup_{i \in \Delta} \mathcal{K}_i$.

A *situation* is a pair $(M, w)$ consisting of a Kripke structure and a world $w$ in $M$. By using situations, we can inductively give semantics to formulas as follows: for primitive propositions $p$,

$$(M, w) \models p \text{ iff } \pi^M(w)(p) = \mathbf{true}.$$

Conjunctions and negations are dealt with in the standard way. Finally,

$(M, w) \models K_i\alpha$ iff for all $w' \in W$ such that $w\mathcal{K}_i^M w'$, we have that $(M, w') \models \alpha$; and

$(M, w) \models C_\Delta\alpha$ iff for all $w' \in W$ such that $w\mathcal{C}_\Delta^M w'$, we have that $(M, w') \models \alpha$.

We say a formula $\alpha$ is satisfiable in Kripke structure $M$ if $(M, w) \models \alpha$ for some possible world $w$ in Kripke structure $M$.

A Kripke structure $M$ is called an S5$_n$ Kripke structure if, for every $i$, $\mathcal{K}_i^M$ is an equivalence relation. A Kripke structure $M$ is called a finite Kripke structure if the set of possible worlds is finite. According to the work of Halpern and Moses (1992), we have the following lemma.

**Lemma 7** *If a formula is satisfiable in an S5$_n$ Kripke structure, then so is in a finite S5$_n$ Kripke structure.*

## 3. Knowledge and Weakest Sufficient Conditions

In our framework, a *knowledge structure* is a simple model of reasoning about knowledge. The advantage of this model is, as will be shown later, that agents' knowledge can be computed via the operation of variable forgetting.

### 3.1 Knowledge Structure

**Definition 8** A *knowledge structure* $\mathcal{F}$ with $n$-agents is a $(n+2)$-tuple $(V, \Gamma, O_1, \cdots, O_n)$ where (1) $V$ is a set of propositional variables; (2) $\Gamma$ is a consistent set of propositional formulas over $V$; and (3) for each agent $i$, $O_i \subseteq V$.

The variables in $O_i$ are called agent $i$'s *observable variables*. An assignment that satisfies $\Gamma$ is called a *state* of knowledge structure $\mathcal{F}$. Given a state $s$ of $\mathcal{F}$, we define *agent $i$'s local state* at state $s$ as $s \cap O_i$. Two knowledge structures are said to be equivalent if they have the same set of propositional variables, the same set of states and, for each agent $i$, the same set of agent $i$'s observable variables.

A pair $(\mathcal{F}, s)$ of knowledge structure $\mathcal{F}$ and a state $s$ of $\mathcal{F}$ is called a *scenario*.

Given a knowledge structure $(V, \Gamma, O_1, \cdots, O_n)$ and a set $\mathcal{V}$ of subsets of $V$, we use $\mathcal{E}_{\mathcal{V}}$ to denote a relation between two assignments $s, s'$ on $V$ satisfying $\Gamma$ such that $(s, s') \in \mathcal{E}_{\mathcal{V}}$ iff there exists a $P \in \mathcal{V}$ with $s \cap P = s' \cap P$. We use $\mathcal{E}_{\mathcal{V}}^*$ to denote the transitive closure of $\mathcal{E}_{\mathcal{V}}$.

Let $\mathcal{V}_{\Delta} = \{O_i \mid i \in \Delta\}$. We then have that $(s, s') \in \mathcal{E}_{\mathcal{V}_{\Delta}}$ iff there exists an $i \in \Delta$ with $s \cap O_i = s' \cap O_i$.

A simple instance of knowledge structure is $\mathcal{F}_0 = (\{p, q\}, \{p \Rightarrow q\}, \{p\}, \{q\})$, where $p, q$ are propositional variables. There are two agents for knowledge structure $\mathcal{F}_0$. Variables $p$ and $q$ are observable to agents 1 and 2, respectively. We have that $\mathcal{V}_{\{1,2\}} = \{\{p\}, \{q\}\}$; and for any two subsets $s$ and $s'$ of $\{p, q\}$ that satisfy $p \Rightarrow q$, we have that $(s, s') \in \mathcal{E}_{\mathcal{V}_{\{1,2\}}}^*$.

We now give the semantics of language $\mathcal{L}_n^C$ based on scenarios.

**Definition 9** The satisfaction relationship $\models$ between a scenario $(\mathcal{F}, s)$ and a formula $\varphi$ is defined by induction on the structure of $\varphi$.

1. For each propositional variable $p$, $(\mathcal{F}, s) \models p$ iff $s \models p$.

2. For any formulas $\alpha$ and $\beta$, $(\mathcal{F}, s) \models \alpha \wedge \beta$ iff $(\mathcal{F}, s) \models \alpha$ and $(\mathcal{F}, s) \models \beta$; and $(\mathcal{F}, s) \models \neg \alpha$ iff not $(\mathcal{F}, s) \models \alpha$.

3. $(\mathcal{F}, s) \models K_i \alpha$ iff for all $s'$ of $\mathcal{F}$ such that $s' \cap O_i = s \cap O_i$, $(\mathcal{F}, s') \models \alpha$.

4. $(\mathcal{F}, s) \models C_{\Delta} \alpha$ iff $(\mathcal{F}, s') \models \alpha$ for all $s'$ of $\mathcal{F}$ such that $(s, s') \in \mathcal{E}_{\mathcal{V}_{\Delta}}^*$.

We say that a proposition formula in $\mathcal{L}(V)$ is *$i$-local* if it is over $O_i$. Clearly, agent $i$ knows an $i$-local formula $\varphi$ in $\mathcal{F}$ iff $\Gamma \models \varphi$.

Let $\mathcal{F} = (V, \Gamma, O_1, \cdots, O_n)$ be a knowledge structure. We say that a formula $\alpha$ is realized in knowledge structure $\mathcal{F}$, if for every state $s$ of $\mathcal{F}$, $(\mathcal{F}, s) \models \alpha$. For convenience, by $\mathcal{F} \models \alpha$, we denote formula $\alpha$ is realized in knowledge structure $\mathcal{F}$.

We conclude this subsection by the following lemma, which will be used in the remains of this paper.

**Lemma 10** *Let $V$ be a finite set of variables, $\mathcal{F} = (V, \Gamma, O_1, \cdots, O_n)$ be a knowledge structure, and $s$ be a state of $\mathcal{F}$. Also suppose that $\Delta \subseteq \{1, \cdots, n\}$, and $\mathcal{V}_{\Delta} = \{O_i \mid i \in \Delta\}$. Then*

1. *for any objective formula $\psi$ (i.e., propositional formula over $V$), $(\mathcal{F}, s) \models \psi$ iff $s \models \psi$;*

2. *for any formula $\gamma \in \Gamma$, $(\mathcal{F}, s) \models \gamma$;*

3. *for any $i$-local formula $\beta$, $(\mathcal{F}, s) \models K_i \beta \Leftrightarrow \beta$;*

4. *for any formula $\beta$, if there exists, for each $i \in \Delta$, an $i$-local formula logically equivalent to $\beta$ under $\Gamma$, then $(\mathcal{F}, s) \models C_\Delta \beta \Leftrightarrow \beta$;*

5. *for any formulas $\alpha_1$ and $\alpha_2$, $(\mathcal{F}, s) \models K_i(\alpha_1 \Rightarrow \alpha_2) \Rightarrow (K_i \alpha_1 \Rightarrow K_i \alpha_2)$;*

6. *for any formulas $\alpha_1$ and $\alpha_2$, $(\mathcal{F}, s) \models C_\Delta(\alpha_1 \Rightarrow \alpha_2) \Rightarrow (C_\Delta \alpha_1 \Rightarrow C_\Delta \alpha_2)$;*

7. *for any formula $\alpha$ and $i \in \Delta$, $(\mathcal{F}, s) \models C_\Delta \alpha \Rightarrow K_i C_\Delta \alpha$.*

**Proof:**

1. The first item of this proposition can be proved by induction on the structure of $\psi$. When $\psi$ is a primitive proposition, it is done by the first item of Definition 9. When $\psi$ is of the form of negation or conjunction, the conclusion also follows immediately by the first item of Definition 9.

2. The second item of this proposition can be proved by the first item and the fact $s$ satisfies $\Gamma$.

3. Given an $i$-local formula $\beta$, it suffices to show $(\mathcal{F}, s) \models K_i \beta$ iff $(\mathcal{F}, s) \models \beta$. By the first item of this proposition, we have that $(\mathcal{F}, s) \models \beta$ iff $s \models \beta$. Moreover, as $\beta$ is $i$-local or over $O_i$, for all assignments $s'$ with $s' \cap O_i = s \cap O_i$, we have that $s' \models \beta$ iff $s \models \beta$. Therefore, we get the following three "iff"s: $(\mathcal{F}, s) \models K_i \beta$ iff, for all state $s'$ of $\mathcal{F}$ with $s' \cap O_i = s \cap O_i$, we have that $(\mathcal{F}, s') \models \beta$ iff, for all state $s'$ of $\mathcal{F}$ with $s' \cap O_i = s \cap O_i$, we have that $s' \models \beta$ iff $s \models \beta$. Thus, $(\mathcal{F}, s) \models K_i \beta$ iff $(\mathcal{F}, s) \models \beta$.

4. Suppose that, for each $i \in \Delta$, there exists an $i$-local formula logically equivalent to $\beta$ under $\Gamma$. We need to show $(\mathcal{F}, s) \models C_\Delta \beta \Leftrightarrow \beta$. First, because $(s, s) \in \mathcal{E}_{\mathcal{V}_\Delta} \subseteq \mathcal{E}_{\mathcal{V}_\Delta}^*$, for all formula $\alpha$, we have that $(\mathcal{F}, s) \models C_\Delta \alpha$ implies $(\mathcal{F}, s) \models \alpha$. Therefore, it suffices to prove that $(\mathcal{F}, s) \models \beta \Rightarrow C_\Delta \beta$. Assume $(\mathcal{F}, s) \models \beta$. To prove that $(\mathcal{F}, s) \models C_\Delta \beta$, we need to show that for every assignment $s'$ such that $(s, s') \in \mathcal{E}_{\mathcal{V}_\Delta}^*$, $(\mathcal{F}, s') \models \beta$. From the definition of $\mathcal{E}_{\mathcal{V}_\Delta}^*$, it suffices to show that for every finite sequence of assignments $s_0, \cdots, s_k$ with $s_0 = s$ and $(s_j, s_{j+1}) \in \mathcal{E}_{\mathcal{V}_\Delta}$ $(0 \le j < k)$, we have that for every $j \le k$, $(\mathcal{F}, s_j) \models \beta$. We show this by induction on $j$. When $j = 0$, the result is clearly true. Assume $(\mathcal{F}, s_j) \models \beta$. Now we prove $(\mathcal{F}, s_{j+1}) \models \beta$. Because $(s_j, s_{j+1}) \in \mathcal{E}_{\mathcal{V}_\Delta}$, there is an $i \in \Delta$ such that $O_i \cap s_j = O_i \cap s_{j+1}$. On the other hand, we have that $s_j \models \beta$ iff $s_{j+1} \models \beta$ because $\beta$ is equivalent under $\Gamma$ to an $i$-local formula. Hence, $(\mathcal{F}, s_{j+1}) \models \beta$ as desired.

5. It suffice to show that if $(\mathcal{F}, s) \models K_i(\alpha_1 \Rightarrow \alpha_2)$ and $(\mathcal{F}, s) \models K_i \alpha_1$, then $(\mathcal{F}, s) \models K_i \alpha_2$. Assume that $(\mathcal{F}, s) \models K_i(\alpha_1 \Rightarrow \alpha_2)$ and $(\mathcal{F}, s) \models K_i \alpha_1$, by item 3 of Definition 9 we get that, for all $s'$ of $\mathcal{F}$ with $s' \cap O_i = s \cap O_i$, we have $(\mathcal{F}, s') \models (\alpha_1 \Rightarrow \alpha_2)$ and $(\mathcal{F}, s') \models \alpha_1$. However, by item 2 of Definition 9, we get $(\mathcal{F}, s') \models \alpha_2$ from $(\mathcal{F}, s') \models (\alpha_1 \Rightarrow \alpha_2)$ and $(\mathcal{F}, s') \models \alpha_1$. Therefore, we get that, for all $s'$ of $\mathcal{F}$ with $s' \cap O_i = s \cap O_i$, we have $(\mathcal{F}, s') \models \alpha_2$. It follows immediately that $(\mathcal{F}, s) \models K_i \alpha_2$.

6. This item can be shown in the same way as in the proof of item 5.

7. It suffices to prove that for those state $s''$ such that there is a state $s'$ with $s \cap O_i = s' \cap O_i$ and $s'\mathcal{E}_{\mathcal{V}_\Delta}^* s''$, we can get $s\mathcal{E}_{\mathcal{V}_\Delta}^* s''$, which follows immediately from the fact that $\mathcal{E}_{\mathcal{V}_\Delta}^*$ is the transitive closure of $\mathcal{E}_{\mathcal{V}_\Delta}$. □

## 3.2 Relationship with S5 Kripke Structure

Given a knowledge structure $\mathcal{F} = (V, \Gamma, O_1, \cdots, O_n)$, let $M(\mathcal{F})$ be Kripke structure $(W, \pi, \mathcal{K}_1, \cdots, \mathcal{K}_n)$, where

1. $W$ is the set of all states of $\mathcal{F}$;

2. for each $w \in W$, the assignment $\pi(w)$ is the same as $w$; and

3. for each agent $i$ and assignments $w, w' \in W$, we have that $w\mathcal{K}_i w'$ iff $w \cap O_i = w' \cap O_i$.

The following proposition indicates that a knowledge structure can be viewed as a specific Kripke structure.

**Proposition 11** *Given a knowledge structure $\mathcal{F}$, a state $s$ of $\mathcal{F}$, and a formula $\alpha$ in $\mathcal{L}_n^C(V)$, we have that $(\mathcal{F}, s) \models \alpha$ iff $(M(\mathcal{F}), s) \models \alpha$.*

**Proof:** Immediately by the definition of the satisfaction relationship between a scenario and a formula and that between a situation and a formula. □

From Proposition 11, we conclude that if a formula in $\mathcal{L}_n^C$ is satisfiable in some knowledge structure, then the formula is also satisfiable in some Kripke structure. From the following proposition and Lemma 7, we can get that if a formula in $\mathcal{L}_n^C$ is satisfiable in some Kripke structure, then the formula is also satisfiable in some knowledge structure.

**Proposition 12** *For a finite $S5_n$ Kripke structure $M$ with the propositional variable set $V$ and possible world $w$ in $M$, there exists a knowledge structure $\mathcal{F}_M$ and a state $s_w$ of $\mathcal{F}$ such that, for every formula $\alpha \in \mathcal{L}_n^C(V)$, we have that $(\mathcal{F}_M, s_w) \models \alpha$ iff $(M, w) \models \alpha$.*

**Proof:** Let $M = (W, \pi, R_1, \cdots, R_n)$, where $W$ is a finite set and $R_1, \cdots, R_n$ are equivalence relations. Let $O_1, \cdots, O_n$ be sets of new propositional variables such that

1. $O_1, \cdots, O_n$ are finite and pairwise disjoint; and

2. for each $i$ $(0 < i \leq n)$, the number of all subsets of $O_i$ is not less than that of all equivalence classes of $R_i$.

By the latter condition, there is, for each $i$, a function $g_i\colon W \mapsto 2^{O_i}$ such that for all $w_1, w_2 \in W$, $g_i(w_1)$ and $g_i(w_2)$ are the same subset of $O_i$ iff $w_1$ and $w_2$ are in the same equivalence class of $R_i$.

Let $V' = V \cup \bigcup_{0 < i \leq n} O_i$. We define a function $g\colon W \mapsto 2^{V'}$ as follows. For each possible world $w$ in $W$,

$$g(w) = \{v \in V \mid \pi(w)(v) = \mathbf{true}\} \cup \bigcup_{0 < i \leq n} g_i(w).$$

The following two claims hold:

C1 For all $w_1, w_2 \in W$, and $i$ $(0 < i \le n)$, we have that $g(w_1) \cap O_i = g(w_2) \cap O_i$ iff $w_1 R_i w_2$.

C2 For all $w \in W$ and $v \in V$, we have that $v \in g(w)$ iff $\pi(w)(v) = \mathbf{true}$.

Let
$$\Gamma_M = \{\alpha \mid \alpha \text{ is over } V', \text{ and } g(w) \models \alpha \text{ for all } w \in W\}.$$

We then get the knowledge structure

$$\mathcal{F}_M = (V', \Gamma_M, O_1, \cdots, O_n).$$

We now show the following claim:

C3 For every $s \subseteq V'$, $s$ is a state of $\mathcal{F}_M$ iff $s = g(w)$ for some $w \in W$.

The "if" part of claim C3 is easy to prove. If $s = g(w')$ for some $w' \in W$, then by the definition of $\Gamma_M$, we have that $g(w') \models \Gamma_M$ and hence $g(w')$ is a state of $\mathcal{F}_M$. To show the "only if" part, assume that for every $w \in W$, $s \neq g(w)$. Then, for every $w \in W$, there exists $\alpha_w$ over $V'$ such that $s \models \alpha_w$ but $g(w) \models \neg \alpha_w$. Therefore, $s \models \bigwedge_{w \in W} \alpha_w$. Moreover, we have that, for every $w' \in W$, $g(w') \models \bigvee_{w \in W} \neg \alpha_w$, and hence $\bigvee_{w \in W} \neg \alpha_w \in \Gamma_M$. Consequently, we have that $s \not\models \Gamma_M$ and hence $s$ is not a state of $\mathcal{F}_M$.

To complete the proof, it suffices to show, for every $\alpha \in \mathcal{L}_n^C(V)$, that $(\mathcal{F}_M, g(w)) \models \alpha$ iff $(M, w) \models \alpha$. With conditions C1, C2 and C3, we can do so by induction on $\alpha$. For the base case, we assume $\alpha$ is a propositional variable, say $p$. Then, by condition C2, we have that $(\mathcal{F}_M, g(w)) \models p$ iff $p \in g(w)$ iff $\pi(w)(p) = \mathbf{true}$ iff $(M, w) \models p$.

Suppose that $\alpha$ is not a propositional variable and the claim holds for every subformula of $\alpha$. There are three cases:

1. $\alpha$ is of form $\neg\beta$ or $\beta \wedge \gamma$. This case can be dealt with by the definitions of satisfaction relations directly.

2. $\alpha$ is of form $K_i\beta$. In this case, we have $(\mathcal{F}_M, g(w)) \models K_i\beta$ iff $(\mathcal{F}_M, s) \models \beta$ for all states $s$ of $\mathcal{F}_M$ with $g(w) \cap O_i = s \cap O_i$. By condition C3, we have that $(\mathcal{F}_M, g(w)) \models K_i\beta$ iff $(\mathcal{F}_M, g(w')) \models \beta$ for all $w' \in W$ with $g(w) \cap O_i = g(w') \cap O_i$. By condition C1, we then have $(\mathcal{F}_M, g(w)) \models K_i\beta$ iff $(\mathcal{F}_M, g(w')) \models \beta$ for all $w' \in W$ with $w R_i w'$. Therefore, by the induction assumption, we have $(\mathcal{F}_M, g(w)) \models K_i\beta$ iff $(M, w') \models \beta$ for all $w' \in W$ with $w R_i w'$. The right part is just $(M, w) \models K_i\beta$.

3. $\alpha$ is of form $C_\Delta\beta$. Recall that, for arbitrary two states $s$ and $s'$ of $\mathcal{F}_M$, $(s, s') \in \mathcal{E}_{\mathcal{V}_\Delta}$ iff there exists an $i \in \Delta$ with $s \cap O_i = s' \cap O_i$. By condition C1, for all $w_1, w_2 \in W$,

$$(g(w_1), g(w_2)) \in \mathcal{E}_{\mathcal{V}_\Delta} \text{ iff } (w_1, w_2) \in \bigcup_{i \in \Delta} R_i.$$

As $\mathcal{E}_{\mathcal{V}_\Delta}^*$ is the transitive closure of $\mathcal{E}_{\mathcal{V}_\Delta}$, and $\mathcal{C}_\Delta^M$ is that of $\bigcup_{i \in \Delta} R_i$, by condition C3 we get that

$$(g(w_1), g(w_2)) \in \mathcal{E}_{\mathcal{V}_\Delta}^* \text{ iff } (w_1, w_2) \in \mathcal{C}_\Delta^M$$

for all $w_1, w_2 \in W$.

We want to show that $(\mathcal{F}_M, g(w)) \models C_\Delta \beta$ iff $(M, w) \models C_\Delta \beta$. On one hand, $(\mathcal{F}_M, g(w)) \models C_\Delta \beta$ iff for all states $s$ of $\mathcal{F}_M$ with $(g(w), s) \in \mathcal{E}_{\mathcal{V}_\Delta}^*$, $(\mathcal{F}_M, s) \models \beta$. By condition C3, we have that $(\mathcal{F}_M, g(w)) \models C_\Delta \beta$ iff for all $w' \in W$ with $(g(w), g(w')) \in \mathcal{E}_{\mathcal{V}_\Delta}^*$. On the other hand, $(M, w) \models C_\Delta \beta$ iff for all $w' \in W$ with $(w, w') \in \mathcal{C}_\Delta^M$. Therefore, we conclude that $(\mathcal{F}_M, g(w)) \models C_\Delta \beta$ iff $(M, w) \models C_\Delta \beta$ by the above discussion. □

Propositions 11 and 12 show that the satisfiability issue for a formula in the language of multi-agent S5 with the common knowledge modality is the same whatever satisfiability is meant w.r.t. a standard Kripke structure or w.r.t. a knowledge structure.

### 3.3 Knowledge as Weakest Sufficient Conditions

The following theorem establishes a bridge between the notion of knowledge and the notion of weakest sufficient and strongest necessary conditions.

**Theorem 13** *Let $V$ be a finite set of variables, $\mathcal{F} = (V, \Gamma, O_1, \cdots, O_n)$ a knowledge structure, $\alpha$ a propositional formula in $\mathcal{L}(V)$, and for an agent $i$, $WSC_i^\alpha$ and $SNC_i^\alpha$ a weakest sufficient condition and a strongest necessary condition of $\alpha$ over $O_i$ under $\Gamma$ respectively. Then, for each state $s$ of $\mathcal{F}$,*

$$(\mathcal{F}, s) \models K_i \alpha \Leftrightarrow WSC_i^\alpha$$

*and*

$$(\mathcal{F}, s) \models \neg K_i \neg \alpha \Leftrightarrow SNC_i^\alpha.$$

**Proof:** We only show $(\mathcal{F}, s) \models K_i \alpha \Leftrightarrow WSC_i^\alpha$, while the other part comes in a straightforward way by duality between WSCs and SNCs. Because $WSC_i^\alpha$ is a sufficient condition of $\alpha$ under $\Gamma$, we have $\Gamma \models WSC_i^\alpha \Rightarrow \alpha$. Let $\theta$ be the conjunction of all formulas in $\Gamma$, then we have $\models \theta \Rightarrow (WSC_i^\alpha \Rightarrow \alpha)$, which leads to $(\mathcal{F}, s) \models K_i WSC_i^\alpha \Rightarrow K_i \alpha$ (by item 5 of Lemma 10.) Because $WSC_i^\alpha$ is $i$-local, by Lemma 10 (item 3) again, we have $(\mathcal{F}, s) \models WSC_i^\alpha \Rightarrow K_i WSC_i^\alpha$. Hence, $(\mathcal{F}, s) \models WSC_i^\alpha \Rightarrow K_i \alpha$.

To show the other direction $(\mathcal{F}, s) \models K_i \alpha \Rightarrow WSC_i^\alpha$, we consider the formula $\forall (V - O_i)(\theta \Rightarrow \alpha)$, where $\theta$ is the same as above. By Proposition 6, we have $\Gamma \models \forall (V - O_i)(\theta \Rightarrow \alpha) \Rightarrow WSC_i^\alpha$. On the other hand, we know that $(\mathcal{F}, s) \models K_i \alpha \Rightarrow \forall (V - O_i)(\theta \Rightarrow \alpha)$ by the definition of $K_i \alpha$. This proves $(\mathcal{F}, s) \models K_i \alpha \Rightarrow WSC_i^\alpha$. □

The following corollary characterizes the subjective formulas $K_i \alpha$ (where $\alpha$ is objective) which are satisfied in a given knowledge structure.

**Corollary 14** *Let $V$ be a finite set of variables, $\mathcal{F} = (V, \{\theta\}, O_1, \cdots, O_n)$ a knowledge structure with $n$ agents, and $\alpha$ a formula over $V$. Then, for every state $s$ of $\mathcal{F}$,*

$$(\mathcal{F}, s) \models K_i \alpha \Leftrightarrow \forall (V - O_i)(\theta \Rightarrow \alpha).$$

**Proof:** Immediately by Theorem 13 and Proposition 6. □

**Example 15:** Now we consider the communication scenario between Alice and Bob addressed in section 1 once again. To show how our system can deal with the knowledge reasoning issue in this scenario, we define a knowledge structure $\mathcal{F}$ as follows:

$$\mathcal{F} = (V, \{\theta\}, O_A, O_B),$$

where

- $O_A = \{Alice\_send\_msg, Alice\_recv\_ack\}$,

- $O_B = \{Bob\_recv\_msg, Bob\_send\_ack\}$,

- $V = O_A \cup O_B$, and

- $\theta$ is the conjunction of the following three formulas:

$$Bob\_recv\_msg \Rightarrow Alice\_send\_msg,$$
$$Bob\_send\_ack \Rightarrow Bob\_recv\_msg,$$
$$Alice\_recv\_ack \Rightarrow Bob\_send\_ack,$$

Now given a state of $\mathcal{F}$

$$s = \left\{ \begin{array}{l} Alice\_send\_msg, \\ Alice\_recv\_ack, \\ Bob\_recv\_msg, \\ Bob\_send\_ack \end{array} \right\},$$

we would like to know whether Alice knows that Bob received the message. Consider the formula

$$\forall \left\{ \begin{array}{l} Bob\_recv\_msg, \\ Bob\_send\_ack \end{array} \right\} (\theta \Rightarrow Bob\_recv\_msg).$$

From Definition 1, the above formula is simplified as $Alice\_recv\_ack$, which, obviously, is satisfied in the scenario $(\mathcal{F}, s)$, i. e. ,

$$(\mathcal{F}, s) \models Alice\_recv\_ack.$$

Then from Corollary 14, we have

$$(\mathcal{F}, s) \models K_A Bob\_recv\_msg.$$

From item 3 of lemma 10, it follows that

$$(\mathcal{F}, s) \models K_A Alice\_send\_msg$$

and

$$(\mathcal{F}, s) \models K_A Alice\_recv\_ack,$$

which indicates that Alice knows that she sent the message and she knows that she received acknowledgement from Bob. $\square$

Given a set of states $S$ of a knowledge structure $\mathcal{F}$ and a formula $\alpha$, by $(\mathcal{F}, S) \models \alpha$, we mean that for each $s \in S$, $(\mathcal{F}, s) \models \alpha$. The following proposition presents an alternative way to compute an agent's knowledge.

**Proposition 16** *Let $V$ be a finite set of variables, $\mathcal{F} = (V, \Gamma, O_1, \cdots, O_n)$ a knowledge structure with $n$ agents, $\psi$ a formula over $V$, and $\alpha$ a formula in $\mathcal{L}_n^C$. Suppose that $SNC_i^\psi$ is a strongest necessary condition of $\psi$ over $O_i$ under $\Gamma$, $S_\psi$ denotes the set of those states $s$ of $\mathcal{F}$ such that $(\mathcal{F}, s) \models \psi$, and $S_{SNC_i^\psi}$ denotes the set of those states $s$ such that $(\mathcal{F}, s) \models SNC_i^\psi$. Then, for each agent $i$, we have that*

$$(\mathcal{F}, S_\psi) \models K_i\alpha \text{ iff } (\mathcal{F}, S_{SNC_i^\psi}) \models \alpha.$$

**Proof:** Let $S_1$ be the set of all states $s$ satisfying $(\mathcal{F}, s) \models \exists(V - O_i)(\theta \wedge \psi)$. Because $\Gamma \models SNC_i^\psi \Leftrightarrow \exists(V - O_i)(\theta \wedge \psi)$, we have $S_1 = S_{SNC_i^\psi}$. Also it is easy to see that for state $s$ of $\mathcal{F}$, $s \in S_1$ iff there is a state $s'$ of $\mathcal{F}$ such that $s' \models \psi$ and $s \cap O_i = s' \cap O_i$. Therefore we have $(\mathcal{F}, S_\psi) \models K_i\alpha$ iff $S_1 \subseteq \{s \mid (\mathcal{F}, s) \models \alpha\}$. This leads to $(\mathcal{F}, S_\psi) \models K_i\alpha$ iff $(\mathcal{F}, S_1) \models \alpha$ iff $(\mathcal{F}, S_{SNC_i^\psi}) \models \alpha$. $\square$

The intuitive meaning behind Proposition 16 is that if all we know about the current state is $\psi$, then all we know about agent $i$'s knowledge (or agent $i$'s observations) is the strongest necessary condition of $\psi$ over $O_i$.

The following proposition provides a method to determined whether a formula with the nested depth of knowledge operators (like $K_{i_1} \cdots K_{i_k}\alpha$, where $\alpha$ is a propositional formula) is always true in those states, where a given proposition formula $\psi$ is true.

**Proposition 17** *Let $V$ be a finite set of variables, $\mathcal{F} = (V, \{\theta\}, O_1, \cdots, O_n)$ a knowledge structure with $n$ agents, $\alpha$ and $\psi$ two formulas over $V$, and $S_\psi$ denotes the set of states $s$ of $\mathcal{F}$ such that $(\mathcal{F}, s) \models \psi$. Then, for each group of agents $i_1, \cdots, i_k$, we have $(\mathcal{F}, S_\psi) \models K_{i_1} \cdots K_{i_k}\alpha$ holds iff*

$$\models \theta \wedge \psi_k \Rightarrow \alpha$$

*where $\psi_k$ is defined inductively as follows:*

$$\psi_1 = \exists(V - O_{i_1})(\theta \wedge \psi);$$

*and for each $j < k$,*

$$\psi_{j+1} = \exists(V - O_{i_{j+1}})(\theta \wedge \psi_j).$$

**Proof:** We show this proposition by induction on the nested depth of knowledge operations. The base case is implied directly by Proposition 16. Assume that the claim holds for those cases with nested depth $k$, we want to show it also holds when the nested depth is $k + 1$, i. e. ,

$$(\mathcal{F}, S_\psi) \models K_{i_1} \cdots K_{i_{k+1}}\alpha \text{ iff } \models \theta \wedge \psi_{k+1} \Rightarrow \alpha.$$

By Proposition 16, we have

$$(\mathcal{F}, S_\psi) \models K_{i_1} \cdots K_{i_{k+1}}\alpha \text{ iff } (\mathcal{F}, S_{\psi_1}) \models K_{i_2} \cdots K_{i_{k+1}}\alpha.$$

By the inductive assumption, we have that

$$(\mathcal{F}, S_{\psi_1}) \models K_{i_2} \cdots K_{i_{k+1}}\alpha \text{ iff } \models \theta \wedge \psi_{k+1} \Rightarrow \alpha.$$

Combining two assertions above, we get

$$(\mathcal{F}, S_\psi) \models K_{i_1} \cdots K_{i_{k+1}} \alpha \text{ iff } \models \theta \wedge \psi_{k+1} \Rightarrow \alpha.$$

□

When we consider the case where the nested depth of knowledge operators is no more than 2, we get the following corollary.

**Corollary 18** *Let* $V, \mathcal{F}, \alpha, \psi$ *and* $S_\psi$ *be as in Proposition 17. Then, for each agent* $i$ *and each agent* $j$, *we have*

1. $(\mathcal{F}, S_\psi) \models K_i \alpha$ *holds iff*

$$\models (\theta \wedge \exists (V - O_i)(\theta \wedge \psi)) \Rightarrow \alpha;$$

2. $(\mathcal{F}, S_\psi) \models K_j K_i \alpha$ *holds iff*

$$\models (\theta \wedge \exists (V - O_i)(\theta \wedge \exists (V - O_j)(\theta \wedge \psi))) \Rightarrow \alpha.$$

**Proof:** Immediately from Proposition 17. □

As will be illustrated in our analysis of security protocols (i.e. Section 6), the part 2 of Corollary 18 is useful for verifying protocol specifications with nested knowledge operators. Given a background knowledge base $\theta$, when we face the task of testing whether $K_j K_i \alpha$ holds in those states satisfying $\psi$, by part 2 of Corollary 18, we can first get $\phi_1 = \exists (V - O_j)(\theta \wedge \psi)$, which is a strongest necessary condition of $\psi$ over $O_j$. This is all we know about what agent $j$ observes from $\psi$. Then we compute $\phi_2 = \exists (V - O_i)(\theta \wedge \phi_1)$, i. e. , the strongest necessary condition of $\phi_1$ over $O_i$ which is, from the viewpoint of agent $j$, about what agent $i$ observes. In this way, the task of checking $K_j K_i \alpha$ is reduced to a task of checking $\theta \wedge \phi_2 \Rightarrow \alpha$.

The following corollary gives two methods to check the truth of $K_i \alpha$ (where $\alpha$ is a propositional formula) in all those states where a given formula $\psi$ is true. One is via the strongest necessary condition of $\psi$ and the other is via the weakest sufficient condition of $\alpha$.

**Corollary 19** *Let* $V$ *be a finite set of propositional variables and* $\mathcal{F} = (V, \{\theta\}, O_1, \cdots, O_n)$ *a knowledge structure with* $n$ *agents,* $\alpha$ *and* $\psi$ *two formulas over* $V$. *Suppose that* $S_\psi$ *denotes the set of all states* $s$ *of* $\mathcal{F}$ *such that* $(\mathcal{F}, s) \models \psi$, *and* $SNC_i^\psi$ *and* $WSC_i^\alpha$ *are a strongest necessary condition of* $\psi$ *over* $O_i$ *and a weakest sufficient condition of* $\alpha$ *over* $O_i$ *under* $\{\theta\}$ *respectively. Then*

1. $(\mathcal{F}, S_\psi) \models K_i \alpha$ *iff* $\models (\theta \wedge \psi) \Rightarrow WSC_i^\alpha$; *and*

2. $(\mathcal{F}, S_\psi) \models K_i \alpha$ *iff* $\models (\theta \wedge SNC_i^\psi) \Rightarrow \alpha$.

**Proof:** The first part of the corollary follows from Theorem 13 and Lemma 10, while the second part follows immediately by Proposition 16. □

In our analysis of security protocols, we observe that very often, it seems more efficient to check an agent's knowledge via the second part of Corollary 19 rather than via the first part. But this may not be always true for some other applications (e.g. see the example of the muddy children puzzle in the next section).

## 4. Common Knowledge

Common knowledge is a special kind of knowledge for a group of agents, which plays an important role in reasoning about knowledge (Fagin et al., 1995). A group of agents $\Delta$ commonly know $\varphi$ when all the agents in $\Delta$ know $\varphi$, they all know that they know $\varphi$, they all know that they all know that they know $\varphi$, and so on ad infinitum. We recall that common knowledge can be characterized in terms of Kripke structures. Given a Kripke structure $M = (W, \pi, \mathcal{K}_1, \cdots, \mathcal{K}_n)$, a group $\Delta$ of agents commonly know $\varphi$ ( or in modal logic language, $C_\Delta \varphi$ is true ) in a world $w$ iff $\varphi$ is true in all worlds $w'$ such that $(w, w') \in \mathcal{C}_\Delta$, where $\mathcal{C}_\Delta$ denotes the transitive closure of $\bigcup_{i \in \Delta} \mathcal{K}_i$.

In this section, we generalize the concept of weakest sufficient and strongest necessary conditions so that they can be used to compute common knowledge.

### 4.1 Generalized Weakest Sufficient and Strongest Necessary Conditions

The following gives a generalized notion of weakest sufficient conditions and strongest necessary conditions.

**Definition 20** Given a set of formulas $\Gamma$ over $V$ as a background knowledge base. Let $\alpha$ be a formula over $V$, and $\mathcal{V}$ a nonempty set of subsets of $V$.

- A formula $\varphi$ is called $\mathcal{V}$-*definable* under $\Gamma$ (or simply called $\mathcal{V}$-*definable* if there is no confusion in the context), if for each $P \in \mathcal{V}$, there is a formula $\psi_P$ over $P$ such that $\Gamma \models \varphi \Leftrightarrow \psi_P$.

- A formula $\varphi$ is called a $\mathcal{V}$-*sufficient condition of* $\alpha$ under $\Gamma$ if it is $\mathcal{V}$-definable and $\Gamma \models \varphi \Rightarrow \alpha$. It is called a *weakest $\mathcal{V}$-sufficient condition of* $\alpha$ under $\Gamma$ if it is a $\mathcal{V}$-sufficient condition of $\alpha$ under $\Gamma$, and for any other $\mathcal{V}$-sufficient condition $\varphi'$ of $\alpha$ under $\Gamma$, we have $\Gamma \models \varphi' \Rightarrow \varphi$.

- Similarly, formula $\varphi$ is called a $\mathcal{V}$-*necessary condition of* $\alpha$ under $\Gamma$ if it is $\mathcal{V}$-definable and $\Gamma \models \alpha \Rightarrow \varphi$. It is called a *strongest $\mathcal{V}$-necessary condition of* $\alpha$ under $\Gamma$ if it is a $\mathcal{V}$-necessary condition of $\alpha$ under $\Gamma$, and for any other $\mathcal{V}$-necessary condition $\varphi'$ of $\alpha$ under $\Gamma$, we have $\Gamma \models \varphi \Rightarrow \varphi'$.

We notice that the notion of $\mathcal{V}$-definability introduced here is a simple elaboration of the notion of V-definability as given in the work of Lang and Marquis (1998): $\phi$ is $\mathcal{V}$-definable under $\Gamma$ iff $\phi$ is V-definable under $\Gamma$ for each $V \in \mathcal{V}$. Moreover, it is easy to see that the formulas implied by $\Gamma$ or inconsistent with it are exactly the formulas $\emptyset$-definable under $\Gamma$, and that definability exhibits a monotonicity property: if $\phi$ is V-definable under $\Gamma$, then $\phi$ is $V'$-definable under $\Gamma$ for each superset $V'$ of $V$ (Lang & Marquis, 1998). Observe also that $\phi$ is V-definable under $\Gamma$ iff $\neg \phi$ is V-definable under $\Gamma$, and this extends trivially to $\mathcal{V}$-definability.

The following lemma says that the notions of weakest $\mathcal{V}$-sufficient conditions and strongest $\mathcal{V}$-necessary ones are dual to each other.

**Lemma 21** *Given a set of formulas $\Gamma$ over $V$ as a background knowledge base, and $\mathcal{V}$ a set of subsets of $V$. Let $\varphi$ and $\alpha$ be formulas over $V$. Then, we have that $\varphi$ is a weakest* $\mathcal{V}$-sufficient condition of $\alpha$ under $\Gamma$ iff $\neg\varphi$ is a strongest $\mathcal{V}$-necessary condition of $\neg\alpha$ under $\Gamma$.

**Proof:**   Straightforward by the duality between WSCs and SNCs. $\square$

To give some intuition and motivation of the above definition, let us consider the following example.

**Example 22:**   Imagine that there are two babies, say Marry and Peter, playing with a dog. Suppose the propositions "The dog is moderately satisfied" (denoted by $m$, for short) and "The dog is full"$(f)$ are understandable to Marry, and the propositions "The dog is hungry" $(h)$ and "The dog is unhappy"$(u)$ are understandable to Peter.

Let $\Gamma = \{h \Rightarrow u, \neg(m \wedge f), (m \vee f) \Leftrightarrow \neg h\}$, $V_1 = \{m, f\}$, $V_2 = \{h, u\}$, and $\mathcal{V} = \{V_1, V_2\}$. We will show that

1. $h$ is $\mathcal{V}$-definable under $\Gamma$;

2. $h$ is a weakest $\mathcal{V}$-sufficient condition of $u$ under $\Gamma$; and

3. $\neg h$ is a strongest $\mathcal{V}$-necessary condition of $\neg u$ under $\Gamma$.

The first claim is easy to check by the definition. The last two claims follow immediately if we can prove that all the $\mathcal{V}$-definable propositions under $\Gamma$ are *false*, *true*, $h$ and $\neg h$ (up to logical equivalence under $\Gamma$). There are 8 propositions over $V_1$ up to logical equivalence. The 8 propositions are: *true*, *false*, $m, \neg m, f, \neg f, m \vee f, \neg m \wedge \neg f$. Similarly, there are 8 propositions over $V_2$ up to logical equivalence under $\Gamma$, i.e., *true*, *false*, $h, \neg h, u, \neg u, h \vee \neg u, \neg h \wedge u$. However, we can find, between the two classes of propositions, only 4 pairs of equivalence relations under $\Gamma$, i.e., $\Gamma \models true \Leftrightarrow true$, $\Gamma \models false \Leftrightarrow false$, $\Gamma \models (m \vee f) \Leftrightarrow \neg h$, $\Gamma \models (\neg m \wedge \neg f) \Leftrightarrow h$. Therefore, all the $\mathcal{V}$-definable propositions under $\Gamma$ are *false*, *true*, $h$ and $\neg h$ (up to logical equivalence under $\Gamma$). $\square$

**Example 23:**   Now we recall the background knowledge $\Gamma_{CS}$ about the communication scenario between Alice and Bob in the introduction section. $\Gamma_{CS}$ is the set of the following three formulas:

$$Bob\_recv\_msg \Rightarrow Alice\_send\_msg$$
$$Bob\_send\_ack \Rightarrow Bob\_recv\_msg$$
$$Alice\_recv\_ack \Rightarrow Bob\_send\_ack$$

Let

$$O_A = \{Alice\_send\_msg, Alice\_recv\_ack\},$$
$$O_B = \{Bob\_recv\_msg, Bob\_send\_ack\},$$
$$\mathcal{V}_{AB} = \{O_A, O_B\}.$$

Clearly, if a formula $\varphi$ is logically implied by $\Gamma_{CS}$ or inconsistent with $\Gamma_{CS}$, then $\varphi$ is $\mathcal{V}_{AB}$-definable under $\Gamma_{CS}$. Moreover, as in Example 22, we are able to check that there are no $\mathcal{V}_{AB}$-definable formulas other than those implied by $\Gamma_{CS}$ or inconsistent with $\Gamma_{CS}$. Therefore, given a formula $\alpha$, a weakest $\mathcal{V}_{AB}$-sufficient condition of $\alpha$ under $\Gamma_{CS}$ is implied by $\Gamma_{CS}$ if $\Gamma_{CS} \models \alpha$, or inconsistent with $\Gamma_{CS}$. $\square$

Let $\Gamma$ be a set of formulas, $V$ a set of propositional variables, and $\mathcal{V}$ a set of subsets of $V$. The following proposition gives the existence of weakest $\mathcal{V}$-sufficient and strongest $\mathcal{V}$-necessary conditions. For a given formula $\alpha$ over $V$, a weakest $\mathcal{V}$-sufficient condition $\phi_1$ of $\alpha$ and a strongest $\mathcal{V}$-necessary condition $\phi_2$ of $\alpha$ can be obtained in the proposition. Indeed, the set of assignments satisfying $\phi_1$ and that of assignments satisfying $\phi_2$ can be given in terms of relation $\mathcal{E}_{\mathcal{V}}$.

**Proposition 24** *Given a finite set $V$ of propositional variables, a set $\Gamma$ of formulas over $V$ as a background knowledge base, a formula $\alpha$ over $V$, and a set $\mathcal{V}$ of subsets of $V$. Denote by $S_{WSC}^{\alpha}$ the set of assignments $s$ over $V$ such that $s \models \Gamma$, and for all assignments $s'$ satisfying $\Gamma$ with $(s, s') \in \mathcal{E}_{\mathcal{V}}^*$, $s' \models \alpha$. Also denote by $S_{SNC}^{\alpha}$ the set of assignments $s$ over $V$ such that $s \models \Gamma$, and there exists an $s'$ such that $s' \models \Gamma$, $s' \models \alpha$ and $(s, s') \in \mathcal{E}_{\mathcal{V}}^*$. Then, the following two points hold.*

- *If a formula is satisfied exactly by those assignments in $S_{WSC}^{\alpha}$, then the formula is a weakest $\mathcal{V}$-sufficient condition of $\alpha$ under $\Gamma$; and*

- *If a formula is satisfied exactly by those assignments in $S_{SNC}^{\alpha}$, then the formula is a strongest $\mathcal{V}$-necessary condition of $\alpha$ under $\Gamma$.*

**Proof:** We first prove the former point, and then show the other by Lemma 21. Let $\phi_1$ be a propositional formula over $V$ such that, for all assignments $s$, $s \models \phi_1$ iff $s \in S_{WSC}^{\alpha}$. Then, for every assignment $s \in S_{WSC}^{\alpha}$, we have $s \models \alpha$ because $(s, s) \in \mathcal{E}_{\mathcal{V}}^*$. Thus, $\phi_1 \models \alpha$.

We remark that for arbitrarily given formula $\varphi$ over $V$ and assignment $s$ over $V$, $s \models \forall (V - P)\varphi$ iff for all assignments $s'$ over $V$ such that $s \cap P = s' \cap P$, we have $s' \models \varphi$.

To prove that $\phi_1$ is $\mathcal{V}$-definable, we show that, for each $P \in \mathcal{V}$, $\phi_1 \models \forall (V-P)\phi_1$, which implies that $\phi_1$ is equivalent to the formula $\forall (V-P)\phi_1$ over $P$. To prove $\phi_1 \models \forall (V-P)\phi_1$, in a semantical way, it suffices to show that, for every assignment $s \in S_{WSC}^{\alpha}$ and $s' \models \Gamma$, if $s \cap P = s' \cap P$, then $s' \in S_{WSC}^{\alpha}$. Let $s$ and $s'$ be given as above and suppose $s \cap P = s' \cap P$. Then, $(s, s') \in \mathcal{E}_{\mathcal{V}}$. Given an assignment $t$ such that $t \models \Gamma$, if $(s', t) \in \mathcal{E}_{\mathcal{V}}^*$, then $(s, t) \in \mathcal{E}_{\mathcal{V}}^*$ by $(s, s') \in \mathcal{E}_{\mathcal{V}}$. Thus, $s' \in S_{WSC}^{\alpha}$. This proves that $\phi_1$ is $\mathcal{V}$-definable.

Now we show that $\phi_1$ is a weakest $\mathcal{V}$-sufficient condition under $\Gamma$. Suppose $\phi$ is a $\mathcal{V}$-definable and sufficient condition of $\alpha$ under $\Gamma$, we want to prove that $\Gamma \models \phi \Rightarrow \phi_1$. The semantical argument of such a proof is as follows. Let $s$ be an assignment with $s \models \Gamma$ and $\phi$, we must show that $s \in S_{WSC}^{\alpha}$, i.e., for every assignment $s'$ with $s' \models \Gamma$ such that $(s, s') \in \mathcal{E}_{\mathcal{V}}^*$, $s' \models \alpha$. Because $\Gamma \models \phi \Rightarrow \alpha$, it suffices to show that $s' \models \phi$. By the condition $(s, s') \in \mathcal{E}_{\mathcal{V}}^*$, there is a finite sequence of assignments $s_0, \cdots, s_k$ such that $s_j \models \Gamma$ with $s_0 = s$ and $s_k = s'$, and for every $j < k$, $(s_j, s_{j+1}) \in \mathcal{E}_{\mathcal{V}}$. By the $\mathcal{V}$-definability of $\phi$, we know that for every $j < k$, $s_j \models \phi$ implies $s_{j+1} \models \phi$. Thus, we have $s' \models \phi$ by induction.

Now we prove the second point of this proposition by Lemma 21. Let $\phi_2$ be a propositional formula over $V$ such that, for all assignments $s$, $s \models \phi_2$ iff $s \in S_{SNC}^{\alpha}$. Let $\theta$ be the conjunction of formulas in $\Gamma$. Then, $s \models \neg \phi_2 \wedge \theta$ iff for all assignments $s'$ with $s' \models \Gamma$ such that $s \mathcal{E}_{\mathcal{V}}^* s'$, we have $s' \models \neg \varphi$. Thus, by the first point of this proposition, we have that $\neg \phi_2 \wedge \theta$ is a weakest $\mathcal{V}$-sufficient condition of $\neg \alpha$. Thus, $\phi_2 \vee \neg \theta$ and hence $\phi_2$ is a strongest $\mathcal{V}$-necessary condition of $\alpha$ according to Lemma 21. $\square$

The above proposition can be thought of as a semantical characterization of weakest $\mathcal{V}$-sufficient and strongest $\mathcal{V}$-necessary conditions.

## 4.2 Characterizations with Least and Greatest Fixed Points

We investigate the computation of the weakest $\mathcal{V}$-sufficient and strongest $\mathcal{V}$-necessary conditions by using the notions of a least and a greatest fixed points of an operator, which is introduced as follows. Let $V$ be a set of propositional variables, and $\Lambda$ be an operator (or a mapping) from the set of propositional formulas over $V$ to the set of propositional formulas over $V$. We say a $\psi$ is a *fixed point* of $\Lambda$, if $\models \Lambda(\psi) \Leftrightarrow \psi$. We say a $\psi_0$ is a *greatest fixed point* of $\Lambda$, if $\psi_0$ is a fixed point of $\Lambda$ and for every fixed point $\psi$ of $\Lambda$, we have $\models \psi \Rightarrow \psi_0$. Clearly, any two greatest fixed points are logically equivalent to each other. Thus, we denote a greatest fixed point of $\Lambda$ by $\mathbf{gfp}Z\Lambda(Z)$. Similarly, we say a $\psi_0$ is a *least fixed point* of $\Lambda$, if $\psi_0$ is a fixed point of $\Lambda$ and for every fixed point $\psi$ of $\Lambda$, we have $\models \psi_0 \Rightarrow \psi$. We denote a least fixed point of $\Lambda$ by $\mathbf{lfp}Z\Lambda(Z)$. We say $\Lambda$ is *monotonic*, if for every two formulas $\psi_1$ and $\psi_2$ such that $\models \psi_1 \Rightarrow \psi_2$, we have $\models \Lambda(\psi_1) \Rightarrow \Lambda(\psi_2)$. For a finite set $V$ of propositional variables if $\Lambda$ is monotonic, then there exists a least fixed point and a greatest fixed point (Tarski, 1955).

**Theorem 25** *Let $V$ be a finite set of variables, $\mathcal{F} = (V, \{\theta\}, O_1, \cdots, O_n)$ a knowledge structure, $\alpha$ a formula over $V$, $\Delta \subseteq \{1, \cdots, n\}$, $\mathcal{V}_\Delta = \{O_i \mid i \in \Delta\}$. Assume that $\Lambda_1$ and $\Lambda_2$ are two operators such that*

$$\Lambda_1(Z) = \bigwedge_{i \in \Delta} \forall(\mathbf{x} - O_i)(\theta \Rightarrow Z)$$

*and*

$$\Lambda_2(Z) = \bigvee_{i \in \Delta} \exists(\mathbf{x} - O_i)(\theta \wedge Z).$$

*Then,*

- *a weakest $\mathcal{V}_\Delta$-sufficient condition of $\alpha$ under $\{\theta\}$ is equivalent to $\mathbf{gfp}\ Z(\alpha \wedge \Lambda_1(Z))$; and*

- *a strongest $\mathcal{V}_\Delta$-necessary condition of $\alpha$ under $\{\theta\}$ is equivalent to $\mathbf{lfp}\ Z(\alpha \vee \Lambda_2(Z))$.*

**Proof:** Let $WSC_\Delta^\alpha$ be a weakest $\mathcal{V}_\Delta$-sufficient condition of $\alpha$ under $\{\theta\}$. Note that the operator $(\alpha \wedge \Lambda_1(Z))$ is monotonic and thus there exists a greatest fixed point of it. Let $\psi_1 = \mathbf{gfp}\ Z(\alpha \wedge \Lambda_1(Z))$. To prove the first point of this theorem, we must show that $\theta \models WSC_\Delta^\alpha \Leftrightarrow \psi_1$.

We first show that $\theta \models WSC_\Delta^\alpha \Rightarrow \psi_1$. For this purpose, we only need to prove

1. $\theta \models WSC_\Delta^\alpha \Rightarrow (\alpha \wedge \Lambda_1(\mathbf{true}))$; and

2. for all formulas $\varphi$ on $V$, if $\theta \models WSC_\Delta^\alpha \Rightarrow \varphi$, then $\theta \models WSC_\Delta^\alpha \Rightarrow (\alpha \wedge \Lambda_1(\varphi))$.

The first point is trivially true because $\Lambda_1(\mathbf{true})$ is equivalent to $\mathbf{true}$ and $WSC_\Delta^\alpha$ is a sufficient condition of $\alpha$ under $\{\theta\}$. To show the second point, suppose $\theta \models WSC_\Delta^\alpha \Rightarrow \varphi$. For $i \in \Delta$, let $\alpha_i$ be the formula over $O_i$ such that $\theta \models WSC_\Delta^\alpha \Leftrightarrow \alpha_i$. Then, $\theta \models \alpha_i \Rightarrow \varphi$. It follows that $\models \alpha_i \Rightarrow (\theta \Rightarrow \varphi)$ and hence $\models \alpha_i \Rightarrow \forall(V - O_i)(\theta \Rightarrow \varphi)$ because $\alpha_i$ does not depend on the variables in $(V - O_i)$. So, we have that, for all $i \in \Delta$, $\theta \models WSC_\Delta^\alpha \Rightarrow \forall(V - O_i)(\theta \Rightarrow \varphi)$. The conclusion of the second point follows immediately.

We now show that $\theta \models \psi_1 \Rightarrow WSC_\Delta^\alpha$, or $\theta \models (\theta \Rightarrow \psi_1) \Rightarrow WSC_\Delta^\alpha$. It suffices to show that $\theta \Rightarrow \psi_1$ is $\mathcal{V}_\Delta$-sufficient condition of $\alpha$ under $\{\theta\}$, that is,

1. $\theta \Rightarrow \psi_1$ is $\mathcal{V}_\Delta$-definable; and

2. $\theta \models (\theta \Rightarrow \psi_1) \Rightarrow \alpha$.

By the fact that $\psi_1$ is a fixed point of the operator $(\alpha \wedge \Lambda_1(Z))$, we have that

$$\models \psi_1 \Rightarrow (\alpha \wedge \bigwedge_{i \in \Delta} \forall(\mathbf{x} - O_i)(\theta \Rightarrow \psi_1)).$$

It follows that $\models \psi_1 \Rightarrow \alpha$, and hence $\theta \models (\theta \Rightarrow \psi_1) \Rightarrow \alpha$. To show the other point, for $i \in \Delta$, we need to prove that $\theta \Rightarrow \psi_1$ is equivalent to a formula over $O_i$. By the above, we have that $\psi_1 \Rightarrow \forall(V - O_i)(\theta \Rightarrow \psi_1)$. It follows that $\theta \models (\theta \Rightarrow \psi_1) \Rightarrow \forall(V - O_i)(\theta \Rightarrow \psi_1)$, and hence

$$\theta \models (\theta \Rightarrow \psi_1) \Leftrightarrow \forall(V - O_i)(\theta \Rightarrow \psi_1)$$

because $\models \forall(V - O_i)(\theta \Rightarrow \psi_1) \Rightarrow (\theta \Rightarrow \psi_1)$ holds trivially. Thus $(\theta \Rightarrow \psi_1)$ is equivalent under $\theta$ to $\forall(V - O_i)(\theta \Rightarrow \psi_1)$, which is over $O_i$. This completes the first point of the conclusion of the theorem.

We now show the second point of this theorem by using the first point and Lemma 21.

Let $SNC_\Delta^\alpha$ be a strongest $\mathcal{V}_\Delta$-necessary condition of $\alpha$ under $\{\theta\}$. By Lemma 21, $\neg SNC_\Delta^\alpha$ is a weakest $\mathcal{V}_\Delta$-sufficient condition of $\neg\alpha$ under $\{\theta\}$. Thus, by the first point of this theorem, $\neg SNC_\Delta^\alpha$ is equivalent to $\mathbf{gfp}\ Z(\neg\alpha \wedge \Lambda_1(Z))$ under $\theta$. Hence, $SNC_\Delta^\alpha$ is equivalent to $\neg\mathbf{gfp}\ Z(\neg\alpha \wedge \Lambda_1(Z))$ under $\theta$. However, $\neg\mathbf{gfp}\ Z(\neg\alpha \wedge \Lambda_1(Z))$ is logically equivalent to $\mathbf{lfp}\ Z(\neg(\neg\alpha \wedge \Lambda_1(\neg Z)))$, which is in turn equivalent to $\mathbf{lfp}\ Z(\alpha \vee \Lambda_2(Z))$. This completes the second point of the theorem. $\square$

### 4.3 Common Knowledge as Weakest $\mathcal{V}$-sufficient Conditions

Given a set $\Delta$ of agents and a family $\mathcal{V}_\Delta$ of observable variable sets of these agents, we investigate the relationship between common knowledge and the weakest $\mathcal{V}_\Delta$-sufficient and strongest $\mathcal{V}_\Delta$-necessary conditions.

**Theorem 26** *Let $V$ be a finite set of variables, $\mathcal{F} = (V, \Gamma, O_1, \cdots, O_n)$ a knowledge structure, $\Delta \subseteq \{1, \cdots, n\}$, $\mathcal{V}_\Delta = \{O_i \mid i \in \Delta\}$, $\alpha$ a formula over $V$, and $WSC_\Delta^\alpha$ and $SNC_\Delta^\alpha$ a weakest $\mathcal{V}_\Delta$-sufficient condition and a strongest $\mathcal{V}_\Delta$-necessary condition of $\alpha$ under $\Gamma$ respectively. Then, for every state $s$ of $\mathcal{F}$,*

$$(\mathcal{F}, s) \models C_\Delta \alpha \Leftrightarrow WSC_\Delta^\alpha$$

*and*

$$(\mathcal{F}, s) \models \neg C_\Delta \neg \alpha \Leftrightarrow SNC_\Delta^\alpha.$$

**Proof:** We only show the first part of this theorem, i.e., $(\mathcal{F}, s) \models C_\Delta \alpha \Leftrightarrow WSC_\Delta^\alpha$, by which and Lemma 21 we can get the other part immediately. Because $WSC_\Delta^\alpha$ is a sufficient condition of $\alpha$, we have that $\Gamma \models WSC_\Delta^\alpha \Rightarrow \alpha$. Let $\theta$ be the conjunction of all formulas in $\Gamma$, we have that $\models \theta \Rightarrow (WSC_\Delta^\alpha \Rightarrow \alpha)$, which leads to $(\mathcal{F}, s) \models C_\Delta WSC_\Delta^\alpha \Rightarrow C_\Delta \alpha$ (by point 6 of Lemma 10). Because $WSC_\Delta^\alpha$ is $\mathcal{V}_\Delta$-definable, we have, by point 4 of Lemma 10, $(\mathcal{F}, s) \models WSC_\Delta^\alpha \Rightarrow C_\Delta WSC_\Delta^\alpha$. Hence, $(\mathcal{F}, s) \models WSC_\Delta^\alpha \Rightarrow C_\Delta \alpha$.

To show the other direction $(\mathcal{F}, s) \models C_\Delta \alpha \Rightarrow WSC_\Delta^\alpha$, we consider the formula $\psi_1$ in the proof of Theorem 25, i.e., the greatest fixed point of the operator

$$\Lambda(Z) = \alpha \wedge \bigwedge_{i \in \Delta} \forall (V - O_i)(\theta \Rightarrow Z).$$

Because we already have $(\mathcal{F}, s) \models \psi_1 \Rightarrow WSC_\Delta^\alpha$ by Theorem 25, it suffices to show $(\mathcal{F}, s) \models C_\Delta \alpha \Rightarrow \psi_1$. Because the greatest fixed point $\psi_1$ of the operator $\Lambda$ can be obtained by a finite iteration of the operator with the starting point $\Lambda(\mathbf{true})$, we only need to prove that

1. $\mathcal{F} \models C_\Delta \alpha \Rightarrow \Lambda(\mathbf{true})$; and

2. for an arbitrary propositional formula $\varphi$ over $V$, if $\mathcal{F} \models C_\Delta \alpha \Rightarrow \varphi$, then $\mathcal{F} \models C_\Delta \alpha \Rightarrow \Lambda(\varphi)$.

The first point is trivially true because $\Lambda(\mathbf{true})$ is equivalent to $\alpha$. To prove the second, suppose $\mathcal{F} \models C_\Delta \alpha \Rightarrow \varphi$. Then, for each $i \in \Delta$, $\mathcal{F} \models K_i(C_\Delta \alpha \Rightarrow \varphi)$. Thus, we have that $\mathcal{F} \models C_\Delta \alpha \Rightarrow K_i \varphi$ by points 5 and 7 of Lemma 10. Hence, $\mathcal{F} \models C_\Delta \alpha \Rightarrow \forall (V - O_i)(\theta \Rightarrow \varphi)$ (by Corollary 14). It follows that $\mathcal{F} \models C_\Delta \alpha \Rightarrow \bigwedge_{i \in \Delta} \forall (V - O_i)(\theta \Rightarrow \varphi)$ and hence $\mathcal{F} \models C_\Delta \alpha \Rightarrow \Lambda(\varphi)$. We thus get $\mathcal{F} \models C_\Delta \alpha \Rightarrow \psi_1$. This completes the proof. □

**Proposition 27** *Given $V$, $\mathcal{F}$, $\Delta$, $\mathcal{V}_\Delta$, $\alpha$ as defined in Theorem 26. Let $\psi$ be a formula over $V$. Assume that a strongest $\mathcal{V}_\Delta$-necessary condition of $\psi$ is $SNC_\Delta^\psi$. Denote by $S_\psi$ the set of those states $s$ of $\mathcal{F}$ such that $(\mathcal{F}, s) \models \psi$, and by $S_{SNC_\Delta^\psi}$ the set of those states $s$ such that $(\mathcal{F}, s) \models SNC_\Delta^\psi$. Then, we have*

$$(\mathcal{F}, S_\psi) \models C_\Delta \alpha \text{ iff } (\mathcal{F}, S_{SNC_\Delta^\psi}) \models \alpha.$$

**Proof:** Let $S_1$ be the set of all states $s$ such that there is a state $s'$ with $s' \models \psi$ and $(s', s) \in \mathcal{V}_\Delta$. We have that $(\mathcal{F}, S_\psi) \models C_\Delta \alpha$ iff for every $s \in S_1$, $(\mathcal{F}, s) \models \alpha$. This leads to $(\mathcal{F}, S_\psi) \models C_\Delta \alpha$ iff $(\mathcal{F}, S_1) \models \alpha$. On the other hand, by Proposition 24, we have that $S_1 = S_{SNC_\Delta^\psi}$. Then the conclusion of the proposition follows immediately. □

Note that, in Proposition 27, if $\alpha$ is a propositional formula, we have that $(\mathcal{F}, S_\psi) \models C_\Delta \alpha$ iff $\Gamma \models SNC_\Delta^\psi \Rightarrow \alpha$. Moreover, by Theorem 26, we have $(\mathcal{F}, S_\psi) \models C_\Delta \alpha$ iff $\Gamma \models \psi \Rightarrow WSC_\Delta^\alpha$, where $WSC_\Delta^\alpha$ is a weakest $\mathcal{V}_\Delta$-sufficient of $\alpha$.

## 5. Adding Public Announcement Operator

There is a recent trend of extending epistemic logic with dynamic operators so that the evolution of knowledge can be expressed. The most basic such extension is public announcement logic (PAL), which is obtained by adding an operator for truthful public announcements. The original version of PAL was proposed by Plaza (1989). In this section, we show that public announcement operator can be conveniently dealt with via our notion of knowledge structure.

## 5.1 Public Announcement Logic

Given a set of agents $A = \{1, \ldots, n\}$ and a set $V$ of propositional variables. The language of public announcement logic $(PAL_n)$ is inductively defined as

$$\varphi ::= p | \neg \varphi | \varphi \wedge \psi | K_i \varphi | C_\Delta \varphi | [\varphi] \psi$$

where $p \in V$, $i \in A$ and $\Delta \subseteq A$.

In other words, $PAL_n$ is obtained from epistemic logic $\mathcal{L}_n^C(V)$ by adding public announcement operator $[\varphi]$ for each formula $\varphi$. Formula $[\varphi]\psi$ means that "after public announcement of $\varphi$, formula $\psi$ is true."

We now give the semantics of public announcement logic under Kripke model. Given a Kripke structure $M = (W, \pi, \mathcal{K}_1, \ldots, \mathcal{K}_n)$, the semantics of the new operators is defined as follows.

$M, w \models [\varphi]\psi$ iff $M, w \models \varphi$ implies $M|_\varphi, w \models \psi$, where $M|_\varphi$ is a Kripke structure such that $M|_\varphi = (W', \pi', \mathcal{K}_1', \ldots, \mathcal{K}_n')$ and

- $W' = \{w \in W | M, w \models \varphi\}$,

- $\pi'(w')(p) = \pi(w')(p)$ for each $w' \in W'$ and each $p \in V$, and

- $\mathcal{K}_i' = \mathcal{K}_i \cap (W' \times W')$ for each $i \in A$.

There are some sentences that become false immediately after the announcement of them. Consider, for example, the sentence '$p$ is true but was not commonly known to be true '. By the announcement of the sentence all agents learn that $p$ and therefore $p$ is commonly known. This can be modelled in public announcement logic by valid formula $[\varphi]\neg C_\Delta p$, where $\varphi = p \wedge \neg C_\Delta p$. To see its validity, let $(M, w)$ be an arbitrary situation. If $M, w \models \varphi$, then $M, w \models p$, which implies that $M|_\varphi, w \models C_\Delta p$, and therefore $M|_\varphi, w \models \neg \varphi$.

## 5.2 Semantics under Knowledge Structure

The semantics of public announcement logic can be conveniently characterized by our notion of knowledge structure. We define the satisfaction relationship $\models$ between a scenario $(\mathcal{F}, s)$ and a formula in $PAL_n$. We need only consider those formulas of the form $[\varphi]\psi$; other cases are the same as in Definition 9.

Let $V$ be a finite set of propositional variables and $\mathcal{F} = (\Gamma, V, O_1, \cdots, O_n)$. The semantics definition for the new operators is as follows. First, let $\mathcal{F}|_\varphi$ be the knowledge structure $(\{\theta\}, V, O_1, \cdots, O_n)$, where $\theta$ is a propositional formula on $V$ such that $(\mathcal{F}, s) \models \varphi$ iff $s$ satisfies $\theta$. As $V$ is a finite set, such a propositional formula $\theta$ always exists.

Then, we set that $(\mathcal{F}, s) \models [\varphi]\psi$ iff $(\mathcal{F}, s) \models \varphi$ implies that $(\mathcal{F}|_\varphi, s) \models \psi$.

We remark that if formula $\varphi$ is equivalent to propositional one $\varphi'$ in knowledge structure $\mathcal{F}$, i.e., $\mathcal{F} \models \varphi \Leftrightarrow \varphi'$ for some propositional formula $\varphi'$, then we can simply define $\mathcal{F}|_\varphi$ as $(\Gamma \cup \{\varphi'\}, V, O_1, \cdots, O_n)$.

The following proposition indicates that the semantics of public announcement logic under knowledge structure coincides with that under Kripke model.

**Proposition 28** *(1) Let $V$ be a finite set of propositional variables and $\mathcal{F} = (\Gamma, V, O_1, \cdots, O_n)$. For every state $s$ of $\mathcal{F}$ and every formula $\alpha \in PAL_n$, we have that $(\mathcal{F}, s) \models \alpha$ iff the situation $(M(\mathcal{F}), s) \models \alpha$. (2) For a finite $S5_n$ Kripke structure $M$ and possible world $w$ in $M$, there is a knowledge structure $\mathcal{F}_M$ and a state $s_w$ of $\mathcal{F}$ such that, for every formula $\alpha \in PAL_n$, we have that $(\mathcal{F}_M, s_w) \models \alpha$ iff $(M, w) \models \alpha$.*

**Proof:** (1) Let us proceed by induction on the structure of formula $\alpha$. We consider only the case that $\alpha$ is of the form $[\varphi]\psi$; other cases are straightforward by the definitions.

By the definition, we have that $(\mathcal{F}, s) \models [\varphi]\psi$ iff $(\mathcal{F}, s) \models \varphi$ implies that $(\mathcal{F}|_\varphi, s) \models \psi$. Thus, by the inductive assumption, we have that $(\mathcal{F}, s) \models [\varphi]\psi$ iff $(M(\mathcal{F}), s) \models \varphi$ implies that $(M(\mathcal{F}|_\varphi), s) \models \psi$. We want to show that $(\mathcal{F}, s) \models [\varphi]\psi$ iff $(M(\mathcal{F}), s) \models [\varphi]\psi$. It suffices to show that $M(\mathcal{F}|_\varphi)$ equals $M(\mathcal{F})|_\varphi$ because $(M(\mathcal{F}), s) \models [\varphi]\psi$ iff $(M(\mathcal{F}), s) \models \varphi$ implies that $(M(\mathcal{F})|_\varphi, s) \models \psi$.

First, the set of possible states of $M(\mathcal{F}|_\varphi)$ equals to the set of those states $s'$ of $\mathcal{F}$ with $(\mathcal{F}, s') \models \varphi$. By the inductive assumption, $(\mathcal{F}, s') \models \varphi$ iff $(M(\mathcal{F}), s') \models \varphi$. Thus, the set of possible states of $M(\mathcal{F}|_\varphi)$ equals to the set of those states $s'$ of $\mathcal{F}$ with $(M(\mathcal{F}), s') \models \varphi$, hence equals to the set of possible states of $M(\mathcal{F})|_\varphi$. Second, we have that for each $s'$ of $\mathcal{F}$ with $(M(\mathcal{F}), s') \models \varphi$, $\pi^{M(\mathcal{F}|_\varphi)}(s') = s'$ and $\pi^{M(\mathcal{F})|_\varphi}(s') = \pi^{M(\mathcal{F})}(s') = s'$. Hence $\pi^{M(\mathcal{F}|_\varphi)} = \pi^{M(\mathcal{F})|_\varphi}$. Finally, for all states $s_1$ and $s_2$ of $\mathcal{F}$ with $(M(\mathcal{F}), s_1) \models \varphi$ and $(M(\mathcal{F}), s_2) \models \varphi$, we have that $(s_1, s_2) \in \mathcal{K}_i^{M(\mathcal{F}|_\varphi)}$ iff $(s_1, s_2) \in \mathcal{K}_i^{M(\mathcal{F})}$ iff $s_1 \cap O_i = s_2 \cap O_i$. Moreover, $(s_1, s_2) \in \mathcal{K}_i^{M(\mathcal{F})|_\varphi}$ iff $s_1 \cap O_i = s_2 \cap O_i$. Therefore, $\mathcal{K}_i^{M(\mathcal{F}|_\varphi)} = \mathcal{K}_i^{M(\mathcal{F})|_\varphi}$. This completes the proof for $M(\mathcal{F}|_\varphi) = M(\mathcal{F})|_\varphi$.

(2) Suppose $M = (W_0, \pi_0, R_1, \cdots, R_n)$, where $W_0$ is a finite set and $R_1, \cdots, R_n$ are equivalence relations. We assume also that the set of propositional variables is $V_0$.

Let $O_1, \cdots, O_n$ be sets of new propositional variables such that

1. $O_1, \cdots, O_n$ are finite and pairwise disjoint; and

2. for each $i$ $(0 < i \leq n)$, the number of all subsets of $O_i$ is not less than that of all equivalence classes of $R_i$.

By the latter condition, there is, for each $i$, a function $g_i \colon W_0 \mapsto 2^{O_i}$ such that for all $w_1, w_2 \in W_0$, $g_i(w_1)$ and $g_i(w_2)$ are the same subset of $O_i$ iff $w_1$ and $w_2$ are in the same equivalence class of $R_i$.

Let $V = V_0 \cup \bigcup_{0 < i \leq n} O_i$. We define a function $g \colon W_0 \mapsto 2^V$ as follows. For each possible world $w$ in $W_0$,

$$g(w) = \{v \in V \mid \pi(w)(v) = \mathbf{true}\} \cup \bigcup_{0 < i \leq n} g_i(w).$$

The following two claims hold:

C1 For all $w_1, w_2 \in W_0$, and $i$ $(0 < i \leq n)$, we have that $g(w_1) \cap O_i = g(w_2) \cap O_i$ iff $w_1 R_i w_2$.

C2 For all $w \in W_0$ and $v \in V_0$, we have that $v \in g(w)$ iff $\pi(w)(v) = \mathbf{true}$.

For any $W \subset W_0$, let

$$\Gamma_W = \{\alpha \mid \alpha \text{ is over } V, \text{ and } g(w) \models \alpha \text{ for all } w \in W\}.$$

We then get a knowledge structure

$$\mathcal{F}_W = (V, \Gamma_W, O_1, \cdots, O_n).$$

We now show that following claim:

C3 For every $s \subseteq V$, $s$ is a state of $\mathcal{F}_W$ iff $s = g(w)$ for some $w \in W$.

The "if" part of claim C3 is easy to prove. If $s = g(w')$ for some $w' \in W$, then by the definition of $\Gamma_W$, we have that $g(w') \models \Gamma_W$ and hence $g(w')$ is a state of $\mathcal{F}_M$. To show the "only if" part, assume that for every $w \in W$, $s \neq g(w)$. Then, for every $w \in W$, there exists $\alpha_w$ over $V$ such that $s \models \alpha_w$ but $g(w) \models \neg\alpha_w$. Therefore, $s \models \bigwedge_{w \in W} \alpha_w$. Moreover, we have that, for every $w' \in W$, $g(w') \models \bigvee_{w \in W} \neg\alpha_w$, and hence $\bigvee_{w \in W} \neg\alpha_w \in \Gamma_W$. Consequently, we have that $s \not\models \Gamma_W$ and hence $s$ is not a state of $\mathcal{F}_W$.

To complete the proof of the second part, it suffices to show, for every $\alpha \in PAL_n$, that $(\mathcal{F}_W, g(w)) \models \alpha$ iff $(M|_W, w) \models \alpha$, where $M|_W$ is a Kripke structure such that $M|_\varphi = (W, \pi, R'_1, \dots, R'_n)$ and

- $\pi(w)(p) = \pi_0(w)(p)$ for each $w \in W$ and each $p \in V_0$, and

- $R'_i = R_i \cap (W' \times W')$ for each $i$ with $0 < i \leq n$.

With claims C1, C2 and C3, we can do so by induction on $\alpha$. Again, we consider only the case that $\alpha$ is of the form $[\varphi]\psi$; other cases can be dealt with in the same way as the proof of Proposition 12.

We first show that knowledge structure $\mathcal{F}_W|\varphi$ is equivalent to $\mathcal{F}_{W'}$, where

$$W' = \{w' \in W \mid M_W, w \models \varphi\}.$$

As the two knowledge structures have the same set $V$ of propositional variables and, for each agent $i$, the same set $O_i$ of observable variables to agent $i$, we need only to prove that they have the same set of states. An assignment $s$ on $V$ is a state of $\mathcal{F}_W|\varphi$ iff $s$ is a state of $\mathcal{F}_W$ and $\mathcal{F}_W, s \models \varphi$. Thus, by claim C3, $s$ is a state of $\mathcal{F}_W|\varphi$ iff $s = g(w')$ for some $w' \in W$ with $\mathcal{F}_W, g(w') \models \varphi$. On the other hand, we have, by claim C3 again, that assignment $s$ is a state of $\mathcal{F}_{W'}$ iff $s = g(w')$ for some $w' \in W'$, i.e., $w' \in W$ and $M_W, w' \models \varphi$. However, by the induction assumption, $\mathcal{F}_W, g(w') \models \varphi$ iff $M_W, w' \models \varphi$. Therefore, knowledge structures $\mathcal{F}_W|\varphi$ and $\mathcal{F}_{W'}$ have the same set of states.

To show $(\mathcal{F}_W, g(w)) \models [\varphi]\psi$ iff $(M|_W, w) \models [\varphi]\psi$, we have, by the induction assumption, that $(\mathcal{F}_W, g(w)) \models \varphi$ iff $(M|_W, w) \models \varphi$. Also, by the claim we just proved above, we have that $(\mathcal{F}_W|_\varphi, g(w)) \models \psi$ iff $(\mathcal{F}_{W'}, g(w)) \models \psi$. By the induction assumption again, $(\mathcal{F}_{W'}, g(w)) \models \psi$ iff $M_{W'}, w \models \psi$. By the definition of $W'$, we have that $M_W|_\varphi, w \models \psi$. Hence, $(\mathcal{F}_W|_\varphi, g(w)) \models \psi$ iff $M_W|_\varphi, w \models \psi$. Therefore, by the semantics of the announcement operators in Kripke structure and knowledge structure, we have that $(\mathcal{F}_W, g(w)) \models [\varphi]\psi$ iff $(M|_W, w) \models [\varphi]\psi$. □

The above proposition is a generalization of Propositions 11 and 12 to $\text{PAL}_n$, which shows that the satisfiability issue for a formula in the language of multi-agent S5 with the announcement operators is the same whatever satisfiability is meant w.r.t. a standard Kripke structure or w.r.t. a knowledge structure.

Notice that, for every formula in $PAL_n$, we can get an equivalent propositional formula. More specifically, we have the following:

**Remark 29** Let $V$ be a finite set of propositional variables and $\mathcal{F} = (\{\theta\}, V, O_1, \cdots, O_n)$. Given a formula $\alpha \in PAL_n$, we define a propositional formula $\lfloor \alpha \rfloor^\theta$ by induction on the structure of $\alpha$:

- If $\alpha$ is a propositional formula, then $\lfloor \alpha \rfloor^\theta = \alpha$.

- $\lfloor \alpha \wedge \beta \rfloor^\theta = \lfloor \alpha \rfloor^\theta \wedge \lfloor \beta \rfloor^\theta$.

- $\lfloor K_i \alpha \rfloor^\theta = \forall (V - O_i)(\theta \Rightarrow \lfloor \alpha \rfloor^\theta)$.

- Let $\Delta \subseteq \{1, \cdots, n\}$, $\mathcal{V}_\Delta = \{O_i \mid i \in \Delta\}$. Then

$$\lfloor C_\Delta \alpha \rfloor^\theta = WSC_\Delta^{\lfloor \alpha \rfloor^\theta}$$

  where $WSC_\Delta^{\lfloor \alpha \rfloor^\theta}$ is a weakest $\mathcal{V}_\Delta$-sufficient condition $\lfloor \alpha \rfloor^\theta$ under $\theta$.

- $\lfloor [\varphi]\alpha \rfloor^\theta = \lfloor \alpha \rfloor^{\theta \wedge \lfloor \varphi \rfloor^\theta}$

Then, for every $\alpha \in PAL_n$, we have that $\mathcal{F} \models \alpha \Leftrightarrow \lfloor \alpha \rfloor^\theta$.

## 6. Complexity Results

We are interested in the following problem: given a knowledge structure $\mathcal{F}$ and a formula $\alpha$ in the language of epistemic logic, whether formula $\alpha$ is realized in structure $\mathcal{F}$. This kind of problem is called the realization problem. In this section, we examine the inherent difficulty of the realization problem in terms of computational complexity. In the general case, this problem is PSPACE-Complete; however, for some interesting subset of the language, it can be reduced to co-NP.

Let $\mathcal{L}$ be some epistemic logic (or language). The realization problem for $\mathcal{L}$ is, given a knowledge structure $\mathcal{F}$ and a formula $\alpha \in \mathcal{L}$, to determine whether $\mathcal{F} \models \alpha$ holds.

The realization problem here is closely related to the model checking problem: given an epistemic formula $\alpha$ and a Kripke structure $M$, to determine whether $M \models \mathcal{F}$. By checking the definition of Kripke structure semantics for epistemic logic, we can see that the model checking problem can be solved in polynomial time (with respect to the input size ($\mid M \mid + \mid \alpha \mid$). We can determine whether a formula $\alpha$ is realized in a knowledge structure $\mathcal{F}$ by first translating knowledge structure $\mathcal{F}$ into a Kripke structure $M$ then checking $M \models \alpha$. However, the resulting algorithm will be exponential in space. This is because the size of the corresponding Kripke structure $M$ is exponential with respect to knowledge structure $\mathcal{F}$.

A number of algorithms for model checking epistemic specifications and the computational complexity of the related realization problems were studied in (van der Meyden, 1998). However, like Kripke structure, the semantics framework they adopt is to list all global states explicitly. As a result, the size of the input of the concerned decision problem can be very large.

**Proposition 30** *The realization problem for $\mathcal{L}_n$ is PSPACE-complete.*

**Proof:** The proposition is of two parts: the PSPACE-easiness and the PSPACE-hardness. The PSPACE-easiness part means that there is an algorithm that determines in polynomial space whether an epistemic formula $\alpha \in \mathcal{L}_n$ is realized in a knowledge structure $\mathcal{F}$. The PSPACE-completeness indicates that there is a PSPACE-hard problem, say the satisfiability problem for quantified propositional formulas (QBF) (Stockmeyer & Meyer, 1973), can be effectively reduced to the realization problem we consider.

It is not difficult to see the PSPACE-easiness. Given a knowledge structure and epistemic formula $\alpha$, by Corollary 14, we can replace knowledge modalities by propositional quantifiers in formula $\alpha$. So, the problem of whether $\alpha$ is realized in $\mathcal{F}$ is reduced to determine whether a quantified Boolean formulas is valid. The latter can be done in polynomial space (Stockmeyer & Meyer, 1973).

As for the PSPACE-hardness, it suffices to show that for every QBF formula

$$\forall p_1 \exists q_2 \forall p_2 \exists q_3 \cdots \forall p_{m-1} \exists q_m A(p_1, q_2, p_2, q_3 \cdots, p_{m-1}, q_m),$$

we can construct a knowledge structure $\mathcal{F}$ such that

$$\vdash \forall p_1 \exists q_2 \forall p_2 \exists q_3 \cdots \forall p_{m-1} \exists q_m A(p_1, q_2, p_2 \cdots, p_{m-1}, q_m)$$

iff

$$\mathcal{F} \models d_1 \wedge \neg d_2 \Rightarrow (K_1 \neg K_2 \neg)^{m-1}(d_m \wedge A(p_1, q_2, p_2, q_3 \cdots, p_{m-1}, q_m)).$$

Let $\mathcal{F} = (V, \{\theta\}, O_1, O_2)$, where

1. $V = \{c\} \cup \{d_1, \cdots, d_m\} \cup \{d'_1, \cdots, d'_m\} \cup \{p_1, \cdots, p_m\} \cup \{q_1, \cdots, q_m\}$

2. $\theta$ is the conjunction of the following formulas

    (a)
    $$\bigwedge_{j < m} (d_{j+1} \Rightarrow d_j) \wedge (d'_{j+1} \Rightarrow d'_j)$$

    (b)
    $$\bigwedge_{j < m} \left( d_j \wedge \neg d_{j+1} \Rightarrow \bigwedge_{i \neq j}(p_i \Leftrightarrow q_i) \right)$$

    (c)
    $$c \Rightarrow \bigwedge_{j < m+1} (d_j \Leftrightarrow d'_j)$$

    (d)
    $$\neg c \Rightarrow \left( ((d_{m-1} \wedge \neg d_m) \Leftrightarrow d'_m) \wedge \bigwedge_{j < m-1} \left( (d_j \wedge \neg d_{j+1}) \Leftrightarrow (d'_{j+1} \wedge \neg d'_{j+2}) \right) \right)$$

3. $O_1 = \{c\} \cup \{d_1, \cdots, d_m\} \cup \{q_1, \cdots, q_m\}$

4. $O_2 = \{d'_1, \cdots, d'_m\} \cup \{p_1, \cdots, p_m\}$

In our picture, we have only two agents: agents 1 and 2. We assign every state an integer number, called the depth of the state for convenience. For every $j$, $d_j$ expresses that the depth of the state is at least $j$. Propositions $d_1, \cdots, d_m$ are observable to agent 1, but not to agent 2. Nevertheless, agent 2 can observe $d'_1, \cdots, d'_m$, which are closely related to $d_1, \cdots, d_m$. The formula in item 2c indicates that $d'_1, \cdots, d'_m$ are the same as $d_1, \cdots, d_m$ if $c$ holds, while the formula in item 2d says that, if $c$ does not hold, the depth expressed by $d_1, \cdots, d_m$ is less than that by $d'_1, \cdots, d'_m$ and the difference is 1. The formula in item 2b implies that, under the condition that the depth of the state is exactly $j$, only $p_j$ is unobservable to agent 1 and only $q_j$ is unobservable to agent 2.

In order to show that

$$\vdash \forall p_1 \exists q_2 \forall p_2 \exists q_3 \cdots \forall p_{m-1} \exists q_m A(p_1, q_2, p_2 \cdots, p_{m-1}, q_m)$$

implies

$$\mathcal{F} \models d_1 \wedge \neg d_2 \Rightarrow (K_1 \neg K_2 \neg)^{m-1}(d_m \wedge A(p_1, q_2, p_2, q_3 \cdots, p_{m-1}, q_m)),$$

it suffices to prove that, for every $j \leq m$ and propositional formula $\varphi$ over $p_1, \cdots, p_m$, $q_1, \cdots, q_m$,

$$\mathcal{F} \models d_j \wedge \neg d_{j+1} \wedge \forall p_j \exists q_{j+1} \varphi \Rightarrow K_1 \neg K_2 \neg (d_{j+1} \wedge \neg d_{j+2} \wedge \varphi)$$

To do so, we need only to show that

$$\mathcal{F} \models d_j \wedge \neg d_{j+1} \wedge \forall p_j \varphi \Rightarrow K_1(d_j \wedge \neg d_{j+1} \wedge \varphi)$$

and

$$\mathcal{F} \models d_j \wedge \neg d_{j+1} \wedge \exists q_{j+1} \varphi \Rightarrow \neg K_2 \neg (d_{j+1} \wedge \neg d_{j+2} \wedge \varphi).$$

As for the other direction, we notice that, for each $l < m - 1$,

$$\mathcal{F} \models d_1 \wedge \neg d_2 \Rightarrow (K_1 K_2)^l \neg d_{l+2}.$$

We also notice that, for each $1 < m' \leq m$,

$$\mathcal{F} \models K_1 \neg K_2 d_{m'} \Rightarrow d_{m'-1}$$

and

$$\mathcal{F} \models d_{m'-1} \wedge \neg d_{m'} \wedge K_1 \neg K_2 \neg (d_{m'} \wedge \varphi) \Rightarrow \forall p_{m'-1} \exists q_{m'} \varphi.$$

By applying the above three claims repeatedly, we can obtain that

$$\mathcal{F} \models d_1 \wedge \neg d_2 \wedge (K_1 \neg K_2 \neg)^{m-1}(d_m \wedge \varphi) \Rightarrow \forall p_1 \exists q_2 \forall p_2 \exists q_3 \cdots \forall p_{m-1} \exists q_m \varphi.$$

Therefore, if

$$\mathcal{F} \models d_1 \wedge \neg d_2 \Rightarrow (K_1 \neg K_2 \neg)^{m-1}(d_m \wedge \varphi)$$

then we have that $\forall p_1 \exists q_2 \forall p_2 \exists q_3 \cdots \forall p_{m-1} \exists q_m \varphi$ is satisfiable in $\mathcal{F}$ because so is $d_1 \wedge \neg d_2$. However, as the QBF formula $\forall p_1 \exists q_2 \forall p_2 \exists q_3 \cdots \forall p_{m-1} \exists q_m \varphi$ does not contain any free variable, we immediately conclude that the QBF formula is valid from that QBF formula is satisfiable in $\mathcal{F}$. $\square$

By Remark 29, we can see that, for the language of formulas in $PAL_n$ without common knowledge operators, the realization problem can be reduced to the problem of validness problem of quantified Boolean formulas, and hence is PSPACE-complete by Proposition 30. We conjecture that the realization problem is also PSPACE-complete for $\mathcal{L}_n^C$ and $PAL_n$.

Proposition 30 indicates that the realization problem in the general case is hard for a computer to solve. Thus, it is interesting to give some special cases with lower computational complexity. Let $\mathcal{L}_n^{+K}$ be the fragment of positive formulas in $\mathcal{L}_n$. It consists of those formulas such that the negation can be applied only to propositional formulas and the modalities are restricted to $K_1, \cdots, K_n$. For instance, formula $K_1 K_2 p \vee K_1 K_2 \neg p$ (where $p$ is a propositional formula) belongs to $\mathcal{L}_n^{+K}$, but formula $K_1 K_2 p \vee K_1 \neg K_2 p$ does not.

The sublanguage $L_n^{+K}$ is interesting in that it is sufficient to represent most important security properties for security protocols. Moreover, as shown in the following proposition, the complexity of the realization problem for $\mathcal{L}_n^{+K}$ is co-NP-complete.

**Proposition 31** *The realization problem for $\mathcal{L}_n^{+K}$ is co-NP-complete.*

**Proof:**    It is well-known that the validity problem for propositional formulas is co-NP-complete. We can easily get the co-NP-hardness of the realization problem for $\mathcal{L}_n^{+K}$, because the validity problem for propositional formulas can be reduced to the realization problem for propositional formulas (considering the case where background knowledge base is a tautology).

On the other hand, to show the realization problem for $\mathcal{L}_n^{+K}$ is in co-NP, we show it can be reduced to the validity problem of propositional formulas. Given a knowledge structure $\mathcal{F}$ and formula $\varphi$ in $\mathcal{L}_n^{+K}$, we will translate $\varphi$ into a propositional formula $\|\varphi\|_{\mathcal{F}}$ (which will be define below), so that $\varphi$ is realized in $\mathcal{F}$ iff $\theta \Rightarrow \|\varphi\|_{\mathcal{F}}$ is valid, where $\theta$ is the background knowledge base of knowledge structure $\mathcal{F}$.

Suppose $\mathcal{F} = (V, \{\theta\}, O_1, \cdots, O_n)$. For every subformula $K_i \psi$ of $\varphi$, we introduce a set $V_\psi^i$ of new propositional variables such that $| V_\psi^i | = | V - O_i |$.

The propositional translation $\|\varphi\|_{\mathcal{F}}$ is inductively given as follows.

1. If $\varphi$ is a propositional formula, then $\|\varphi\|_{\mathcal{F}} = \varphi$.

2. If $\varphi$ is of the conjunction form $\varphi_1 \wedge \varphi_2$, then

$$\|\varphi\|_{\mathcal{F}} = \|\varphi_1\|_{\mathcal{F}} \wedge \|\varphi_2\|_{\mathcal{F}}.$$

3. If $\varphi$ is of the disjunction form $\varphi_1 \vee \varphi_2$, then

$$\|\varphi\|_{\mathcal{F}} = \|\varphi_1\|_{\mathcal{F}} \vee \|\varphi_2\|_{\mathcal{F}}.$$

4. If $\varphi$ is of the form $K_i \psi$, then

$$\|\varphi\|_{\mathcal{F}} = (\theta \Rightarrow \|\psi\|_{\mathcal{F}})(\frac{V - O_i}{V_\psi^i}),$$

where $(\theta \Rightarrow \|\psi\|_{\mathcal{F}})(\frac{V - O_i}{V_\psi^i})$ is the formula obtained from $(\theta \Rightarrow \|\psi\|_{\mathcal{F}})$ by replacing variables in $V - O_i$ by the new ones in $V_\psi^i$.

The idea behind the above translation is that we first translate formula $\varphi$ into a quantified propositional formula, where all the quantifiers are universal ones, and then eliminate those universal quantifiers by introducing new variables.

Let $V_\varphi$ be the set of new variables in $\|\varphi\|_\mathcal{F}$. To show the correctness of the translation, it suffices to show that $\mathcal{F} \models \varphi \Leftrightarrow \forall V_\varphi \|\varphi\|_\mathcal{F}$.

We prove this claim by induction on $\varphi$.

- It is trivial, if $\varphi$ is a propositional formula.

- If $\varphi$ is of the form $\varphi_1 \wedge \varphi_2$, the claim can be obtained immediately by the induction assumption.

- If $\varphi$ is of the form $\varphi_1 \vee \varphi_2$, we have that $\forall V_\varphi(\|\varphi_1\|_\mathcal{F} \vee \|\varphi_2\|_\mathcal{F})$ is logically equivalent to $\forall V_{\varphi_1}\|\varphi_1\|_\mathcal{F} \vee \forall V_{\varphi_2}\|\varphi_2\|_\mathcal{F}$, as the variables in $V_{\varphi_1}$ do not appear in $\forall V_{\varphi_2}\|\varphi_2\|_\mathcal{F}$ and those in $V_{\varphi_2}$ do not in $\forall V_{\varphi_1}\|\varphi_1\|_\mathcal{F}$. Thus, the claim holds by the induction assumption.

- Finally, if $\varphi$ is of the form $K_i\psi$, then

$$\|\varphi\|_\mathcal{F} = (\theta \Rightarrow \|\psi\|_\mathcal{F})(\frac{V - O_i}{V_\psi^i}).$$

Therefore, $V_\varphi = V_\psi \cup V_\psi^i$ and $\forall V_\varphi \|\varphi\|_\mathcal{F}$ is logically equivalent to $(\theta \Rightarrow \forall V_\psi^i \forall V_\psi \|\psi\|_\mathcal{F})(\frac{V-O_i}{V_\psi^i})$. Thus, by the induction assumption, we have that

$$\mathcal{F} \models \forall V_\varphi \|\varphi\|_\mathcal{F} \Leftrightarrow (\theta \Rightarrow \forall V_\psi^i \psi(\frac{V - O_i}{V_\psi^i})$$

and hence

$$\mathcal{F} \models \forall V_\varphi \|\varphi\|_\mathcal{F} \Leftrightarrow (\theta \Rightarrow \forall(V - O_i)\psi).$$

Therefore, we have $\mathcal{F} \models \forall V_\varphi \|\varphi\|_\mathcal{F} \Leftrightarrow K_i\psi$. □

Proposition 31 implies that, for an arbitrary formula $\varphi$ in $\mathcal{L}_n^{+K}$ and a knowledge structure $\mathcal{F}$ with background knowledge base $\theta$,

$$\mathcal{F} \models \varphi \text{ iff } \theta \wedge \neg\|\varphi\|_\mathcal{F} \text{ is unsatisfiable.}$$

Thus, we can solve the realization problem for formulas in $\mathcal{L}_n^{+K}$ by using a propositional satisfiability solver.

## 7. A Case Study: the Muddy Children Puzzle

In this section, we demonstrate how our framework can be applied to practical problems by using the example of the muddy children puzzle.

### 7.1 Muddy Children Puzzle

The muddy children puzzle is a well-known variant of the wise men puzzle. The story goes as follows (Fagin et al., 1995): Imagine $n$ children playing together. Some of the children, say $k$ of them, get mud on their foreheads. Each can see the mud on others but not on his/her own forehead. Along comes the father, who says, "at least one of you has mud on your forehead." The father then asks the following question, over and over: "Does any of you know whether you have mud on your own forehead?"

Assuming that all children are perceptive, intelligent, truthful, and they answer simultaneously, what we want to show is that the first $(k-1)$ times the father asks the question, they will say "No" but the $k^{th}$ time the children with muddy foreheads will all answer "Yes."

### 7.2 Modeling the Muddy Children Puzzle

To model the muddy children puzzle, let $m_i$ be a propositional variable, which means that child $i$ is muddy $(i < n)$. Denote by $V$ the set $\{m_i \mid i < n\}$. Suppose the assignment $s_0 = \{m_i \mid i < k\}$ represents the actual state: child $0$, $\cdots$, child $k-1$ have mud on their foreheads; and the other children have not. This can be captured by the scenario $(\mathcal{F}_0, s_0)$, where $\mathcal{F}_0 = (V, \Gamma_0, O_0, \cdots, O_{n-1})$ with

- $V = \{m_i \mid i < n\}$;

- $\Gamma_0 = \emptyset$;

- and $O_i = V - \{m_i\}$ for each $i < n$.

Let $\varphi = \bigwedge_{i<n} \neg K_i m_i$, which indicates that every child does not know whether he has mud on his own forehead. For convenience, we introduce, for all natural number $l$, the notations $[\varphi]^l \psi$ so that $[\varphi]^0 \psi = \psi$ and $[\varphi]^{l+1} \psi = [\varphi][\varphi]^l \psi$. The properties we want to show is then formally expressed in $PAL_n$:

- $[\bigvee_{i<n} m_i][\varphi]^j \varphi$ for every $0 \leq j < k-1$, and

- $[\bigvee_{i<n} m_i][\varphi]^{k-1} \bigwedge_{i<k} K_i m_i$.

Formula $[\bigvee_{i<n} m_i][\varphi]^j \varphi$ means that the children will all say "No" for the $j+1^{th}$ time the father asks the question. In particular, when $j = 0$, the condition $0 \leq j < k-1$ is simplified as $k > 1$; and the resulting formula $[\bigvee_{i<n} m_i]\varphi$ says that after the father announces $\bigvee_{i<n} m_i$ every child says "No". Formula $[\bigvee_{i<n} m_i][\varphi]^{k-1} \bigwedge_{i<k} K_i m_i$ indicates that the $k^{th}$ time the children with muddy foreheads will all answer "Yes."

Therefore, what we want to prove is that

$$(\mathcal{F}_0, s_0) \models \left( \bigwedge_{0 \leq j < k-1} [\bigvee_{i<n} m_i][\varphi]^j \varphi \right) \wedge \left( [\bigvee_{i<n} m_i][\varphi]^{k-1} \bigwedge_{i<k} K_i m_i \right).$$

To check the above, we basically follow the definition of $PAL$ semantics under knowledge structure. During the checking process, a series $\mathcal{F}_j$ $(0 < j \leq k)$ of knowledge structures are constructed so that $\mathcal{F}_1 = \mathcal{F}_0 \mid_{\bigvee_{i<n} m_i}$ and, for every $j$ $(0 < j < k)$, $\mathcal{F}_{j+1} = \mathcal{F}_j \mid_{\varphi}$.

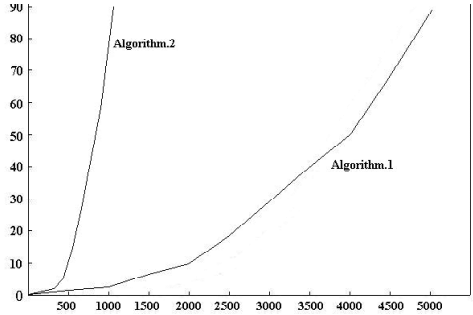

Figure 1: Performances of the two algorithms for the muddy children puzzle

Specifically, we have that, for each step $j \leq k$, we get

$$\mathcal{F}_j = (V, \Gamma_j, O_0, \cdots, O_{n-1})$$

where $O_i = V - \{m_i\}$ for each $i < n$, and $\Gamma_j$ is defined as follows:

- At step 1: $\Gamma_1 = \{\bigvee_{i<n} m_i\}$.

- At step $j + 1$: Let $\varphi^b = \bigwedge_{i<n} \neg \forall m_i(\Gamma_j \Rightarrow m_i)$. As for each $i < n$, $\mathcal{F}_j \models K_i m_i \Leftrightarrow \forall m_i(\Gamma_j \Rightarrow m_i)$, we have that $\mathcal{F}_j \models \varphi \Leftrightarrow \varphi^b$. Thus, we may set $\Gamma_{j+1} = \Gamma_j \cup \{\varphi^b\}$.

Therefore, it suffices to verify, for $0 < j < k$ and $i < n$, $(\mathcal{F}_j, s_0) \models \neg K_i m_i$, and for $i < k$, $(\mathcal{F}_k, s_0) \models K_i m_i$.

## 7.3 Experimental Results

Our framework of knowledge structure has been implemented by using the BDD library (CUDD) developed by Fabio Somenzi at Colorado University. Notice that BDD-based QBF solvers for satisfiability problems are not among the best solvers nowadays. However, in the experiments here we need to compute and represent a serial of Boolean functions (say $\Gamma_j$), which are not decision problems and can not be solved by a general QBF solver.

To check agents' knowledge, we implemented two different algorithms in terms of Part 1 and 2 of Corollary 19 in Section 3, respectively. Algorithm 1, which is based on part 1 of Corollary 19, seems much more efficient than Algorithm 2, which is based on part 2 of Corollary 19, for this particular example. The reason is as follows. It is clear that the main task of both algorithms is to check whether $(\mathcal{F}_j, s_0) \models K_i m_i$. However, Algorithm 1's method is to compute $s_0 \models \forall m_i(\Gamma_j \Rightarrow m_i)$, while Algorithm 2 is to compute $\models \exists m_i(\Gamma_j \wedge s_0) \Rightarrow m_i$. Now the main reason why Algorithm 1 is much more efficient for this particular problem is clear: $\forall m_i(\Gamma_j \Rightarrow m_i)$ is simply equivalent to $\neg \Gamma_j(\frac{m_i}{false})$. Assuming half of the children are muddy, Fig. 1 gives the performances for a Pentium IV PC at 2.4GHz, with 512RAM. In the figure, the x-axis is for the number of children, and the y-axis for the CPU run time in seconds.

The muddy children puzzle as a famous benchmark problem of reasoning about knowledge can be resolved by both proof-theoretic and semantical approaches (Baltag et al., 1998; Gerbrandy, 1999; Lomuscio, 1999). Proof-theoretic approaches depend on efficient provers for multi-modal logics; and semantical ones may suffer from the state-explosion problem. Our approach is essentially a semantic one, but we give a syntactical and compact way to represent Kripke structures by using knowledge structures, and hence may avoid the state-explosion problem to some extent.

## 8. Application to Verification of Security Protocols

In this section, we apply our knowledge model to security protocol verification. Security protocols that set up credits of the parties and deal with the distribution of cryptographic keys are essential in communication over vulnerable networks. Authentication plays a key role in security protocols. Subtle bugs that lead to attack are often found when the protocols have been used for many years. This presents a challenge of how to prove the correctness of a security protocol. Formal methods are introduced to establish and prove whether a secure protocol satisfies a certain authentication specification.

### 8.1 Background on Authentication Protocols

Authentication protocols aim to coordinate the activity of different parties (usually referred to as *principals*) over a network. They generally consist of a *sequence* of message exchanges whose format is fixed in advance and must be conformed to. Usually, a principal can take part into a protocol run in different ways, as the *initiator* or the *responder*; we often call the principal has different *roles*. Very often a principal can take part into several protocol runs simultaneously with different roles.

The designers of authentication protocols must have the conscious in mind that the message may be intercepted and someone with malicious intention can impersonate an honest principal. One of the key issues in authentication is to ensure the *confidentiality*, that is, to prevent private information from being disclosed to unauthorized entities. Another issue is to avoid intruder impersonating other principals. In general, a principal should ensure that the message he receives was created *recently* and sent by the principal who claims to have sent it.

Cryptography is a fundamental element in authentication. A message transmitted over a channel without any cryptographic converting is called *plaintext*. The intention of cryptography is to transform a given message to some form that is unrecognizable by anyone except the intended receiver. The procedure is called *encryption* and the corresponding parameter is known as *encryption key*. The encoded message is referred to as *ciphertext*. The reverse procedure is called *decryption* and uses the corresponding *decryption key*. The *symmetric-key cryptography*, which is also called *secret-key cryptography*, uses the same key for both encryption and decryption. The *asymmetric-key cryptography*, which is also called *public-key cryptography*, uses different keys for encryption and decryption. The one for the encryption is the *public key* that is generally available for anyone. Corresponding to the public key is the *private key*, which is for the decryption and only owned by one principal.

## 8.2 The Dolev-Yao Intruder Model

The standard adversary model for the analysis of security protocols was introduced by Dolev and Yao in 1983 and is commonly known as *Dolev-Yao model* (Dolev & Yao, 1983). According to this model, a set of conservative assumptions are made as follows:

1. Messages are considered as indivisible abstract values instead of sequences of bits.

2. All the messages from one principal to any other principals must pass through the adversary and the adversary acts as a general router in the communication.

3. The adversary can read, alter and redirect any message.

4. The adversary can only decrypt a message if he has the right keys, and can only compose new messages from keys and messages that he already possesses.

5. The adversary can not perform any statistical or other cryptanalytic attacks.

Although this model has the drawback of finding implementation dependent attacks, it simplifies the protocol analysis. It has been proved to be the most powerful modeling of the adversary (Cervesato, 2001) because it can simulate any other possible attackers.

## 8.3 The Revised Needham-Schroeder Protocol

As Lowe (1996) pointed out, the Needham-Schroeder protocol has the problem of lacking the identity of the responder and can be fixed by a small modification. However, it is not clear if the revised version is correct. Our approach provides a method to automatically prove the correctness of security protocols instead of just finding bugs as usual analysis tools do for security protocols.

In the cryptography literature, the revised Needham-Schroeder protocol is described as follows:

1. $A \rightarrow B$: $\{Na, A\}_{Kb}$

2. $B \rightarrow A$: $\{B, Na, Nb\}_{Ka}$

3. $A \rightarrow B$: $\{Nb\}_{Kb}$

where $A \rightarrow B : M$ is a notation for "$A$ sends $B$ the message $M$" or "$B$ receives the message $M$ from $A$". The notation $\{M\}_K$ means the encryption of $M$ with the key $K$. Also, $A, B$ denote the principal identifiers; and $Ka, Kb$ indicate, respectively, $A$'s and $B$'s public keys. Moreover, $Na$ and $Nb$ are the *nonces* which are newly generated unguessable values by A and B, respectively, to guarantee the freshness of messages.

Two informal goals or specifications of the protocol are "$A$ knows that $B$ knows $A$ said $Na$ and $Na$ is fresh," and "$B$ knows that $A$ knows $B$ said $Nb$ and $Nb$ is fresh ."

To analyze the protocol, we introduce $A$ and $B$ *local histories* for the protocol: If $A$ plays the role of the initiator in the protocol, and assumes that $B$ be the responsor, then $A$'s local history is that

1. $A$ said $\{Na, A\}_{Kb^A}$

2. $A$ sees $\{B^A, Na, Nb^A\}_{Ka}$

3. $A$ said $\{Nb^A\}_{Kb^A}$

where "A said $M$" means that $A$ sent the message $M$, or other message containing $M$; "A sees $M$" indicates that $A$ receives $M$ or got $M$ by some received messages; $B^A$ is the responsor of the protocol from $A$'s local view; $Kb^A$ and $Nb^A$ are, from $A$'s local view, the responsor's public key and nonce, respectively.

If $B$ plays the role of the responsor in the protocol, and assumes $A$ be the initiator, then $A$'s local history is that

1. $B$ sees $\{Na^B, A^B\}_{Kb}$

2. $B$ said $\{B, Na^B, Nb\}_{Ka}$

3. $B$ sees $\{Nb\}_{Kb}$

where $A^B$ is the initiator of the protocol from $B$'s local observations; $Ka^B$ and $Na^B$ are, from $B$'s local view, the initiator's public key and nonce, respectively.

The main point of our analysis is that if an agent is involved in the protocol, then the agent's real observations should be compatible with the so-called *local history*. For example, if $A$ is the initiator of the protocol, and $A$ sees $\{B, Na^B, Nb\}_{Ka}$, then according to $A$'s local history for the protocol we have that $A$ assumes that $B$ is the responsor of the protocol, the responsor's nonce is $Nb$, and from the responsor's view, the initiator's nonce is $Na$ (see the 4th formula of the background knowledge $\Gamma$ below).

Let us see how our framework of reasoning about knowledge can be applied to this protocol.

The variable set $V_{RNS}$ consists of the following atoms:

- $fresh(Na)$: Nonce $Na$ is fresh.

- $fresh(Nb)$: Nonce $Nb$ is fresh.

- $role(Init, A)$: $A$ plays the role of the initiator of the protocol.

- $role(Resp, B)$: $B$ plays the role of the responder of the protocol.

- $Resp^A = B$: $A$ assumes that the responder of the protocol is $B$.

- $Init^B = A$: $B$ assumes that the initiator of the protocol is $A$.

- $Na^B = Na$: $B$ assumes that the partner's nonce in the execution of the protocol is $Na$.

- $Nb^A = Nb$: $A$ assumes that the partner's nonce in the execution of the protocol is $Nb$.

- $said(B, Na)$: $B$ said $Na$ by sending a message containing $Na$.

- $said(A, Nb)$: $A$ said $Nb$.

- $sees(B, \{Na, A\}_{Kb})$: $B$ sees $\{Na, A\}_{Kb}$ (possibly by decrypting the messages received.)

- $sees(A, \{B, Na^B, Nb\}_{Ka})$: $A$ sees $\{B, Na^B, Nb\}_{Ka}$.

The background knowledge $\Gamma_{RNS}$ consists of the following formulas:

1. $\begin{pmatrix} sees(B, \{Na, A\}_{Kb}) \wedge \\ said(B, Na) \wedge \\ fresh(Na) \end{pmatrix} \Rightarrow role(Resp, B)$

2. $\begin{pmatrix} sees(A, \{B, Na^B, Nb\}_{Ka}) \wedge \\ said(A, Nb) \wedge \\ fresh(Nb) \end{pmatrix} \Rightarrow role(Init, A)$

3. $\begin{pmatrix} role(Resp, B) \wedge \\ sees(B, \{Na, A\}_{Kb}) \wedge \\ said(B, Na) \wedge \\ fresh(Na) \end{pmatrix} \Rightarrow \begin{pmatrix} Init^B = A \wedge \\ Na^B = Na \end{pmatrix}$

4. $\begin{pmatrix} role(Init, A) \wedge \\ sees(A, \{B, Na^B, Nb\}_{Ka}) \wedge \\ said(A, Nb) \wedge \\ fresh(Nb) \end{pmatrix} \Rightarrow \begin{pmatrix} Resp^A = B \wedge \\ Na^B = Na \wedge \\ Nb^A = Nb \end{pmatrix}$

5. $\begin{pmatrix} role(Init, A) \wedge \\ Resp^A = B \end{pmatrix} \Rightarrow \begin{pmatrix} sees(B, \{Na, A\}_{Kb}) \wedge \\ said(B, Na) \end{pmatrix}$

6. $\begin{pmatrix} role(Resp, B) \wedge \\ Init^B = A \end{pmatrix} \Rightarrow \begin{pmatrix} sees(A, \{B, Na^B, Nb\}_{Ka}) \wedge \\ said(A, Nb) \end{pmatrix}$

7. $(role(Init, A) \Rightarrow fresh(Na)) \wedge$
   $(role(Resp, B) \Rightarrow fresh(Nb))$

Notice that the first two formulas are required for the rationality of the agents $A$ and $B$. The other formulas in $\Gamma$ can be obtained automatically by some fixed set of meta rules. We obtain the third and fourth formulas by comparing their local history for the protocols to the conditions appearing in the formulas. To get the fifth formula informally, consider $A$'s local history under the conditions $role(Init, A)$ and $Resp^A = B$, which should be that

1. $A$ said $\{Na, A\}_{Kb}$

2. $A$ sees $\{B, Na, Nb^A\}_{Ka}$

3. $A$ said $\{Nb^A\}_{Kb}$.

According to $A$'s local history, $A$ sees the nonce $Na$ generated by $A$ itself. Because $Na$ is only said in the message $\{Na, A\}_{Kb}$, thus $B$, who has the inverse key of $Kb$, must see this message and said $Na$. Similarly, we can see that the sixth formula holds. The last formula follows immediately by the definition of the protocol.

The set $O_A$ of the observable variables to $A$ is

$$\{fresh(Na), role(Init, A), Resp^A = B\}.$$

The set $O_B$ of the observable variables to $B$ is

$$\{fresh(Nb), role(Resp, B), Init^B = A\}.$$

Now consider the knowledge structure

$$\mathcal{F} = (V_{RNS}, \Gamma_{RNS}, O_A, O_B).$$

Let $Spec_A$ be the formal specification:

$$\begin{pmatrix} fresh(Na) \wedge \\ role(Init, A) \wedge \\ Resp^A = B \end{pmatrix} \Rightarrow K_A K_B \begin{pmatrix} said(A, Na) \wedge \\ fresh(Na) \end{pmatrix}$$

and $Spec_B$ be the formal specification:

$$\begin{pmatrix} fresh(Nb) \wedge \\ role(Resp, B) \wedge \\ Init^B = A \end{pmatrix} \Rightarrow K_B K_A \begin{pmatrix} said(B, Nb) \wedge \\ fresh(Nb) \end{pmatrix}.$$

It is easy to show that, for all states $s$ of $\mathcal{F}$,

$$(\mathcal{F}, s) \models Spec_A \wedge Spec_B$$

as desired.

We should mention that, in the original Needham-Schroeder protocol (Needham & Schroeder, 1978), the second message is $B \rightarrow A$: $\{Na, Nb\}_{Ka}$ instead of $B \rightarrow A$: $\{B, Na, Nb\}_{Ka}$. Therefore, the fourth formula in $\Gamma$ would be changed to

$$\begin{pmatrix} role(Init, A) \wedge \\ sees(A, \{Na^B, Nb\}_{Ka}) \wedge \\ said(A, Nb) \wedge \\ fresh(Nb) \end{pmatrix} \Rightarrow \begin{pmatrix} Na^B = Na \wedge \\ Nb^A = Nb \end{pmatrix}$$

Thus, $Resp^A = B$ does not necessarily hold under the condition

$$role(Init, A) \wedge sees(A, \{Na^B, Nb\}_{Ka}) \wedge said(A, Nb) \wedge fresh(Nb).$$

This is why the specifications $Spec_A$ and $Spec_B$ do not hold for the original Needham-Schroeder protocol.

## 8.4 Discussion

BAN logic (Burrows, Abadi, & Needham, 1990) is one of the most successful logical tools to reason about security protocols. However, the semantics of BAN is always arguable, and it is not clear under what assumption the rules of BAN logic is sound and complete. This motivated the research of seeking more adequate frameworks (models). Providing a model-theoretic semantics for BAN logic has been a central idea in the development of BAN-like logics such as AT (Abadi & Tuttle, 1991) and SVO (Syversion & van Oorschot, 1996). The advantage of our approach is that we use knowledge structures as semantic models to verify the correctness of epistemic goals for security protocols.

An important problem is that, given a security protocol, where and how the corresponding knowledge structure comes from. To get the knowledge structure corresponding to a security protocol, we have developed a semantic model, and the background knowledge base of the corresponding knowledge structure consists of those formulas valid in the semantic model. Moreover, we can generate the background knowledge systematically. The ongoing work is to implement our approach into a promising automatic security protocol verifier.

## 9. Related Work

There are a number of approaches dealing with the concept of variable forgetting or *eliminations of middle terms* (Boole, 1854) in several contexts. The notion of variable forgetting was formally defined in propositional and first order logics by Lin and Reiter (1994). In recent years, theories of forgetting under answer set programming semantics were proposed (Zhang & Foo, 2006; Eiter & Wang, 2008). Forgetting was also generalized to description logics (Kontchakov, Wolter, & Zakharyaschev, 2008; Wang, Wang, Topor, & Pan, 2008; Kontchakov, Walther, & Wolter, 2009).

In the context of epistemic logic, the notion of forgetting was studied in a number of ways. Baral and Zhang (2006) treated knowledge forgetting as a special form of update with the effect $\neg K\varphi \wedge \neg K\neg\varphi$: after knowledge forgetting $\varphi$, the agent would neither know $\varphi$ nor $\neg\varphi$. Ditmarsch, Herzig, Lang and Marquis (2008) proposed a dynamic epistemic logic with an epistemic operator $K$ and a dynamic modal operator $[Fg(p)]$ so that formula $[Fg(p)]\varphi$ means that after the agent forgets his knowledge about $p$, $\varphi$ is true. (Zhang & Zhou, 2008) modeled forgetting via bisimulation invariance except for the forgotten variable. This notion of variable forgetting is closely related to quantified modal logics, where the existential variable quantification can be modeled via bisimulation invariance except for the quantified variable (Engelhardt et al., 2003).

The notion of *variable forgetting* has various applications in knowledge representation and reasoning. For example, Weber (1986) applied it to updating propositional knowledge bases. Lang and Marquis (2002) used it for merging a set of knowledge bases when simply taking their union may result in inconsistency. The notion of variable forgetting is also closely related to that of *formula-variable independence*, because the result of forgetting the set of variables $V$ in a formula $\varphi$ can be defined as the strongest consequence of $\varphi$ being independent from $V$ (Lang et al., 2003). More recently, Liu and Lakemeyer (2009) applied the notion of forgetting into the situation calculus, and obtained some interesting results about the first-order definability and computability of progression for local-effect actions.

## 10. Conclusion

The main contribution of this paper is as follows. First, we have investigated knowledge reasoning within a simple framework called knowledge structure, which consists of a global knowledge base and a set of observable variables for each agent. The notion of knowledge structure can be used as a semantic model for a multi-agent logic of knowledge and common knowledge. In this model, the computation of knowledge and common knowledge can be reduced to the operation of variable forgetting; moreover, an objective formula $\alpha$ is known by agent $i$ at state $s$ when any of its weakest sufficient condition on $O_i$ holds at state $s$.

Second, to capture the notion of common knowledge in our framework, we have generalized the notion of weakest sufficient conditions and obtained, for a set $\mathcal{V}$ of sets of propositional variables, the notion of the weakest $\mathcal{V}$-sufficient conditions. Given a set $\Delta$ of agents and a family $\mathcal{V}_\Delta$ of observable variable sets of these agents, we have shown that an objective formula $\alpha$ is common knowledge for agents in $\Delta$ iff the weakest $\{O_i \mid i \in \Delta\}$-sufficient of $\alpha$ holds. Also, we have shown that public announcement operator can be conveniently dealt with via our notion of knowledge structure.

Third, the relationship between S5 Kripke structure and knowledge structure has been explored. Specifically, the satisfiability issue for a formula in the language of multi-agent S5 with public announcement operator is the same as what satisfiability is meant w.r.t. a standard Kripke structure or w.r.t. a knowledge structure.

Fourth, we have examined the computational complexity of the problem whether a formula $\alpha$ is realized in structure $\mathcal{F}$. In the general case, this problem is PSPACE-hard; however, there are some interesting subcases in which it can be reduced to co-NP.

Finally, we have shown the strength of the concept of knowledge structure from the practical side by some empirical results about the satisfiability problem for knowledge structures based on the instances of the muddy children puzzle, since even for the smallest instances considered in the experiments generating the corresponding S5 Kripke structure would be out of reach. we have also discussed the automated analysis and verification of the corrected Needham-Schroeder protocol via knowledge structures.

Our work presented in this paper can be further extended in several directions. First, we will investigate whether our knowledge structures can be extended and used as a basis for knowledge based programming (Fagin et al., 1995). Secondly, in our current framework of knowledge structures, we have not considered the issue of *only knowing* which has been extensively studied in other knowledge reasoning models (Halpern & Lakemeyer, 1996; van der Hock, Jaspars, & Thijsse, 2003; Levesque, 1990). It will be an interesting topic of how our knowledge model handles only knowing in reasoning about knowledge. Thirdly, the notions and methods in this work can be extended to investigate the extension of the variable forgetting operator to multi-agent logics of beliefs. Finally, recent research has shown that *knowledge update* has many important applications in reasoning about actions and plans and dynamic modeling of multi-agent systems (Zhang, 2003). A first step in this direction (in mono-agent S5) can be found in the work of Herzig, Lang and Marquis (2003). Baral and Zhang have proposed a general model for performing knowledge update based on the standard single agent S5 modal logic (Baral & Zhang, 2001). We believe that their work can be extended to multi-agent modal logics by using the knowledge structure defined in this paper and therefore to develop a more general system for knowledge update. Along this direction, an interesting research issue is to explore the underlying relationship between *knowledge forgetting* - a specific type of knowledge update, and variable forgetting as addressed in this paper.

## Acknowledgments

The authors thank Ron van der Meyden, Fangzheng Lin and the anonymous reviewers for their valuable comments on an earlier version of this paper. This work was partially supported by the Australian Research Council grant DP0452628, the National Basic Research 973 Program grants (Nos. 2010CB328103, 2009CB320701 and 2005CB321902), and National Natural Science Foundation of China grants (Nos. 60725207 and 60763004). This paper is the revised and extended version of a paper which appeared in *Proceedings of KR 2004* (Su, Lv, & Zhang, 2004)